**Theoretical Study of Molecular Electronic and Rotational Coherences by High-Harmonic Generation**

Song Bin Zhang, [1,2 *] Denitsa Baykusheva, [3] Peter M. Kraus, [3] Hans Jakob Wörner, [3]

and Nina Rohringer [1,2 +]

[1] Max-Planck Institute for the Physics of Complex Systems, 01187 Dresden, Germany

[2] Center for Free-Electron Laser Science (CFEL), DESY, 22607 Hamburg, Germany

[3] Laboratory for physical chemistry, ETH Zürich Wolfgang Pauli Str. 10, 8093 Zürich, Switzerland

## Abstract

The detection of electron motion and electronic wavepacket dynamics is one of the core goals of attosecond science. Recently, choosing the nitric oxide (NO) molecule as an example, we have introduced and demonstrated a new experimental approach to measure coupled valence electronic and rotational wavepackets using high-harmonic generation (HHG) spectroscopy [Kraus *et al.*, Phys. Rev. Lett. **111**, 243005 (2013)]. A short outline of the theory to describe the combination of the pump and HHG probe process was published together with an extensive discussion of experimental results [Baykusheva *et al.*, Faraday Discuss **171**, 113 (2014)]. The comparison of theory and experiment showed good agreement on a quantitative level. Here, we present the generalized theory in detail, which is based on a generalized density matrix approach that describes the pump process and the subsequent probing of the wavepackets by a semiclassical quantitative rescattering approach. An in-depth analysis of the different Raman scattering contributions to the creation of the coupled rotational and electronic spin-orbit wavepackets is made. We present results for parallel and perpendicular linear polarizations of the pump and probe laser pulses. Furthermore, an analysis of the combined rotational-electronic density matrix in terms of irreducible components is presented, that facilitates interpretation of the results.

## I. Introduction

* song-bin.zhang@pks.mpg.de

+ nina.rohringer@pks.mpg.de

Measuring and controlling electronic and nuclear motion is one of the core interests of ultrafast atomic, molecular and optical physics. High Harmonic Generation (HHG) is a sensitive tool to measure the coherent rotational [1-9] and vibrational [10, 11] wavepacket dynamics [12] of a prepared molecular ensemble. Although the detection of coherent electronic wavepackets through the HHG process has been proposed in theory [13-21], it hitherto remained experimentally unexplored. Several experiments have studied the electronic dynamics by HHG spectroscopy [22-26], but the effect of the electronic coherence has not been studied so far. Recently, we demonstrated that the HHG process can map out electronic coherences with high sensitivity [8]. The basic principle of the detection process is a cross channel of the HHG process, that coherently connects different electronic states (see Fig 1a).

In our experiment [8], a supersonically cooled NO molecular beam is firstly irradiated by an IR laser pulse, that creates rotational wavepackets by Raman scattering and prepares the molecular ensemble in a superposition of two electronic states - the $F_1$ ($\Pi_{1/2}$ dominated) and $F_2$ ($\Pi_{3/2}$ dominated) states; In a second step, the rotational and spin-orbit electronic wavepackets are probed by a short laser pulse inducing the process of HHG. HHG spectra are recorded as a function of the delay time between the two pulses. The total HHG yield shows a strong dependence on the delay time. In addition to the known traces of phasing of the rotational wavepackets to an aligned ensemble of molecules, fast oscillations in the HHG yield are observed, that reflect the electronic spin-orbit wavepacket. This trace is inherently connected to the electronic coherence of the wavepacket. Fig. 1a illustrates the concept of the HHG from a coherent superposition of two electronic states: the left two channels depict the conventional HHG pathways – ionization from and recombination to the same state (direct channels); In case of a fixed phase relation between the two electronic states – in terms an electronic density matrix this means a non-zero off-diagonal matrix element - a cross channel contributes to the HHG signal, that coherently connects ionization from one and recombination to the other electronic state.

Several groups worked on theoretical frameworks to describe the process of laser alignment and a subsequent HHG probe process, based on a density matrix formalism of the rigid-rotor states [27-30] or S-matrix theory [31-33] to describe the molecular rotational wavepackets. In these approaches, the HHG process is treated within the single-active electron approximation and the strong-field approximation [34], or the Keldysh-Faisal-Reiss approximation (KFR) [35]. The quantitative rescattering method (QRS) [36-38] is another successful theory for both diatomic and polyatomic molecules that expresses the HHG spectra as the product of a returning electron wavepacket and the photo-recombination cross section. These approaches, however, have not been developed to describe the dynamics of a coupled rotational-electronic coherent wavepacket. Recently we extended these theories to a combined electronic and rotational density-matrix approach, that allows to study the new cross channel of the HHG process [39]. Here, we give an in-depth derivation of this generalized density-matrix formalism of HHG. We analyze the density matrix in terms of irreducible tensor components and also present more detailed calculations that underline the importance of different Raman scattering terms in the interaction Hamiltonian of the pump process.

The theoretical model, including a discussion of the rotational level structure of NO, the description of the Raman pump process and the approximation to determine the HHG spectra are given in section II, followed by a discussion of the results in section III. Finally, we will summarize the work in section IV. We opted for a self-consistent presentation of the theory that necessarily leads to the introduction of some results already presented in Ref. [8, 39].

## II. Methods

### II.A Rotational structure of NO

The nitrogen monoxide (NO) molecule possesses a degenerate $^2\Pi$ open shell electronic ground state. Its field-free effective rotational Hamiltonian is given by the sum of the kinetic energy associated with the rotational motion of the molecule and the spin-orbit interaction term [40, 41]:

$$H_0 = H_{rot} + H_{so} = B(\mathbf{J} - \mathbf{L} - \mathbf{S})^2 + A(\mathbf{L} \cdot \mathbf{S})\,. \quad (1)$$

Here $\boldsymbol{J}$=$\boldsymbol{L}$+$\boldsymbol{S}$+$\boldsymbol{R}$ denotes the total angular momentum of the molecule, $\boldsymbol{L}$ is the total orbital angular momentum, $\boldsymbol{S}$ the total spin angular momentum and $\boldsymbol{R}$ is the angular momentum of the rotating nuclei. The ground state rotational constant and the spin-orbit coupling constant are $B$ = 1.6961 cm$^{-1}$ and $A$ = 123.13 cm$^{-1}$, respectively [42]. The projections of $\boldsymbol{L}$ and $\boldsymbol{S}$ onto the internuclear axis are denoted by $\Lambda$ and $\Sigma$, respectively. $\Omega=\Lambda+\Sigma$, is defined as the projection of $\boldsymbol{J}$ onto the internuclear axis. The Hamiltonian of Eq. (1) can be easily diagonalized by first expressing it in the basis of states of Hund's coupling case (a) $|J,|\Omega|,M_0,\varepsilon\rangle$. Here $M_0$ denotes the projection of the total angular momentum ***J*** onto a laboratory fixed axis. We choose this axis along the polarization direction of the pump-laser field, which creates the rotational wavepackets. $\varepsilon=\pm1$ is a symmetry index, related to the total rotational parity $p$ ($p=\varepsilon(-1)^{J-1/2}$) [40]. Diagonalization of the Hamiltonian of Eq. (1) in that basis then yields the eigenvectors $\{|\Xi JM_0\varepsilon\rangle,\Xi=1,2\}$ and eigenvalues $\{E_J^{(\Xi)},\Xi=1,2\}$ given by [43]

$$\begin{pmatrix}|\Xi=1JM_0\varepsilon\rangle\\|\Xi=2JM_0\varepsilon\rangle\end{pmatrix}=\begin{pmatrix}\mathrm{a}_J & \mathrm{b}_J\\ \text{-}\mathrm{b}_J & \mathrm{a}_J\end{pmatrix}\begin{pmatrix}\left|J,\frac{1}{2},M_0,\varepsilon\right\rangle\\\left|J,\frac{3}{2},M_0,\varepsilon\right\rangle\end{pmatrix}, \tag{2}$$

and

$$\begin{pmatrix}E_J^{(1)}\\E_J^{(2)}\end{pmatrix}=\begin{pmatrix}B[(J-1/2)(J+3/2)-X_J/2]\\B[(J-1/2)(J+3/2)+X_J/2]\end{pmatrix}. \tag{3}$$

The kets $|J,\overline{\Omega},M_0,\varepsilon\rangle$ are defined by

$$|J,\overline{\Omega},M_0,\varepsilon\rangle=\frac{1}{\sqrt{2}}(|J,\overline{\Omega},M_0\rangle+\varepsilon|J,-\overline{\Omega},M_0\rangle), \tag{4}$$

and $|J,\Omega,M_0\rangle$ relates to the Wigner D-matrix $D^J_{M_0,\Omega}$ as

$$\langle\varphi,\theta,\chi\,|\,J,\pm\overline{\Omega},M_0\rangle=\sqrt{\frac{2J+1}{8\pi^2}}D^J_{M_0,\pm\overline{\Omega}}(\varphi,\theta,\chi). \tag{5}$$

Here we defined $\overline{\Omega}=|\Omega|=\frac{1}{2},\frac{3}{2}$, and $(\varphi,\theta,\chi)$ are the Euler angles defining the orientation of the body-fixed frame or molecular frame with respect to the laboratory frame [44]. The coefficients of $\mathrm{a}_J$ and $b_J$ are given by $\mathrm{a}_J=\sqrt{\frac{X_J+Y-2}{2X_J}}$ and $b_J=\sqrt{\frac{X_J-Y+2}{2X_J}}$, where $X_J=\sqrt{4(J+1/2)^2+Y(Y-4)}$ and $Y=A/B$. Generally, the states characterized by Eq. (4) are referred to as $\Pi_{\overline{\Omega}}$, alluding to the total angular momentum projection $\overline{\Omega}=|\Omega|=\frac{1}{2},\frac{3}{2}$ of the spin-orbit coupled electronic state. The eigenstates (described by Eq. (2) and Eq. (3)) are usually referred to as $F_1$ and $F_2$ states. Hund's coupling case (a) is valid for small angular momentum numbers *J*, or when the condition *BJ*<<*A* is satisfied. In that case $a_J\sim1$, $b_J\sim0$, so that the lower $F_1$ state is $\Pi_{1/2}$ dominated, and the upper $F_2$ state is $\Pi_{3/2}$ dominated. The level diagram of the rotational states for the two electronic states are shown in Fig. 1b. It should be noted that, generally, a linear molecule has only two rotational degrees of freedom. The dependence of Eq. (5) on the Euler angle $\chi$ has the form of a mere phase factor $e^{\mp i\bar{\Omega}\chi}$, and for an eigen state, $\chi$ is usually fixed arbitrarily [43, 44]. The choice $\chi = 0$ is conventionally made, so that the molecule's *x* axis lies in the plane spanned by that of the spaced-fixed *Z* axis and the molecule's *z* axis. For eigenstates this convention has to be adopted for both rotational and electronic wave functions. Since we are interested in combined rotational/electronic wavepackets, the angular momentum projections on the molecular axis will generally not have a single value, *i.e.*, the molecular ensemble will be described by fractional occupations of the $\bar{\Omega} = 1/2$ and $\bar{\Omega} = 3/2$ states. The shape of the electronic density of such superposition will generally not be symmetric under rotations around the molecular axis. Therefore, it is necessary to keep the dependence on the angle $\chi$ and, contrary to the usual treatment of eigenstates, do not set it equal to zero. The density matrix, describing the combined rotational/electronic wavepackets will hence depend on the angle $\chi$, as discussed in the following sections.

### II.B The pump process: impulsive laser alignment and creation of electronic / rotational wavepackets

The ensemble of molecules is prepared in a coherent superposition of eigenstates $\{|\Xi JM_0\varepsilon\rangle, \Xi = 1,2\}$ by interaction with an optical laser pulse. The combined rotational and electronic spin-orbit wavepacket is created by optical Raman scattering – the process typically inducing impulsive field-free alignment of the molecular ensemble – induced by the linearly polarized pump laser field $\boldsymbol{\varepsilon}_1(t) = \varepsilon_1(t)\cos(\omega_0 t)\hat{\varepsilon}_1 = \varepsilon_{1,0} e^{-2\ln 2(t/\tau_p)^2} \cos(\omega_0 t)\hat{\varepsilon}_1$, where $\hat{\varepsilon}_1$ is a unit vector parallel to the polarization axis of the pump field. $\varepsilon_{1,0}$ is the electric field amplitude, that is in the range 0.029-0.041 a.u. in the experiment, corresponding to experimental peak intensities in the range of 3-6×$10^{13}$ W/cm$^2$, $\tau_p$ denotes the pulse duration at full width half maximum of the field intensity. In the experiment, $\tau_p$ is estimated to be 60 fs. $\omega_0 = 1.5$ eV is the fundamental frequency of the pump field of 800 nm applied in the experiment. The effective rotational Hamiltonian including the coupling to the linearly polarized laser field can be written as

$H_1(t) = H_0 + H_{\text{int}}(t)$, (6)

where $H_0$ denotes the field-free Hamiltonian of Eq. (1). The effective cycle-averaged field interaction Hamiltonian $H_{\text{int}}(t)$ is given in terms of the polarizability tensor $\underline{\underline{\alpha}}$ and reads [39, 40, 45]

$$\begin{aligned} H_{\text{int}}(t) &= -\frac{\varepsilon_1^2(t)}{4}\hat{\varepsilon}_1 \underline{\underline{\alpha}} \hat{\varepsilon}_1 \\ &= -\frac{\varepsilon_1^2(t)}{2}\frac{1}{\sqrt{6}}[D_{0,0}^2(\varphi,\theta,\chi)T_0^2(\alpha) + (D_{0,2}^2(\varphi,\theta,\chi) + D_{0,-2}^2(\varphi,\theta,\chi))T_2^2(\alpha)]. \\ &= -\frac{\varepsilon_1^2(t)}{4}\frac{2}{3}\Delta\alpha[D_{0,0}^2(\varphi,\theta,\chi) + \gamma(D_{0,2}^2(\varphi,\theta,\chi) + D_{0,-2}^2(\varphi,\theta,\chi))] \end{aligned} \quad (7)$$

Here, we expressed the effective interaction Hamiltonian in terms of the Wigner *D*-matrix, and the polarizability tensor in terms of its irreducible tensor components $T(\alpha)$. The isotropic interaction, proportional to the tensor $T_0^0(\alpha)$ is excluded since it only introduces an overall phase to the total wavefunction. The matrix elements of the interaction Hamiltonian can be easily calculated in the basis set $\left\{\left|J,\overline{\Omega},M_0,\varepsilon\right\rangle\right\}$ (that strictly speaking is not the eigenbasis of the stationary Hamiltonian $H_0$) and selection rules for the transition matrix elements can be given. In the basis set $\left\{\left|J,\overline{\Omega},M_0,\varepsilon\right\rangle\right\}$, the interaction proportional to the terms $D_{0,0}^2(\varphi,\theta,\chi)$ and $D_{0,\pm2}^2(\varphi,\theta,\chi)$ have only nonzero matrix elements for $\Delta\Omega = 0$ and $\Delta\Omega = \pm 2$, respectively. The quadrupolar term proportional to $D_{0,0}^2(\varphi,\theta,\chi)T_0^2(\alpha)$ ($T_0^2(\alpha) = \frac{1}{\sqrt{6}}[2\alpha_{zz} - \alpha_{xx} - \alpha_{yy}]$) therefore predominantly excites higher angular momentum states *J*, i.e., a rotational wavepacket, within the electronic subspace $\Xi = 1$ or $\Xi = 2$. The selection rules for the transition operator $\left\langle J_1,\overline{\Omega_1},M_0,\varepsilon\right| D_{0,0}^2 \left|J_2,\overline{\Omega_2},M_0,\varepsilon\right\rangle$ are $\overline{\Omega_1} = \overline{\Omega_2}$ and $\Delta J = J_2 - J_1 = 0,\ \pm1,\ \pm2$. Since the representation of the interaction Hamiltonian in terms of the eigenbasis of $H_0$ $\{\left|\Xi J M_0 \varepsilon\right\rangle, \Xi = 1,2\}$ has a small contribution in the off-diagonal block connecting states $F_1$ and $F_2$, $D_{0,0}^2$ can induce transitions between different electronic states. The quadrupolar interaction $D_{0,\pm2}^2(\varphi,\theta,\chi)T_2^2(\alpha)$ ($T_2^2(\alpha) = \frac{1}{2}[(\alpha_{xx} - \alpha_{yy}) + 2i\alpha_{xy}]$) is mainly responsible for creating the electronic excitations. The selection rules for the transition operator $\left\langle J_1,\overline{\Omega_1},M_0,\varepsilon\right| D_{0,\pm2}^2 \left|J_2,\overline{\Omega_2},M_0,\varepsilon\right\rangle$ are $\Omega_2 - \Omega_1 = \pm 2$ and $\Delta J = J_2 - J_1 = 0,\ \pm1,\ \pm2$. This means that this interaction term mediates excitations to the other electronic state along with excitations of a rotational wavepacket in the excited electronic state. The relative importance of these two contributions $D^2_{0,0}$ and $D^2_{0,\pm2}$ will be discussed in section III. Here the components of the polarizability tensor are defined in the molecular-fixed Cartesian frame. For the NO molecule the value of the polarizabilities are given by $\alpha_{\|} = \alpha_{zz} = 15.34$ a.u. and $\alpha_{\perp} = \alpha_{xx} = \alpha_{yy}$ =9.715 a.u. [46]. $\Delta\alpha = \alpha_{\|} - \alpha_{\perp} = 5.625$ a.u. is the difference between the parallel and perpendicular components of the polarizability. The parameter $\gamma = < T_2^2(\alpha) / T_0^2(\alpha) >$ quantifies the intensity ratio between electronic and purely rotational Raman scattering and has been attributed the empirical value 0.2 [40].

Initially the molecular ensemble is assumed to be in thermal equilibrium and in the electronic ground state or $F_1$ state. According to the experimental conditions [8], we assume an initial rotational temperature of about T= 10 K of the molecular ensemble and that initially only the $F_1$ state is occupied, so that the initial density matrix is described by a thermal, diagonal density matrix

$$\hat{\rho}(t=0) = \sum_{J_0 M_0} w_{J_0} \sum_{\varepsilon} \frac{1}{2} \left| \Xi = 1 J_0 M_0 \varepsilon \right\rangle \left\langle \Xi = 1 J_0 M_0 \varepsilon \right| , \tag{8}$$

where $w_{J_0}$ are the statistical weights according to a Boltzmann distribution of the rotational degrees of freedom. Fig. 2 gives the initial occupation of the density matrix for the case $T$=10 K. Only states up to $J_{max}$ = 19/2 are considerably occupied, implying that Hund's case (a) is an appropriate description [47]. The density matrix at later times $t$ can be constructed in terms of states $\left| \varphi^{J_0 M_0}(t) \right\rangle$, that follow the evolution under the total Hamiltonian of Eq. (6) for an initial condition $\left| \varphi^{J_0 M_0}(0) \right\rangle = \sum_{\varepsilon} \frac{1}{\sqrt{2}} \left| \Xi = 1 J_0 M_0 \varepsilon \right\rangle$. Note that the quantum number $M_0$ is conserved under the interaction with the linearly polarized field. In practice, we can expand the time-dependent rotational wavefunction $\left| \varphi^{J_0 M_0}(t) \right\rangle$ in terms of the eigen basis $\{ \left| \Xi J M_0 \varepsilon \right\rangle, \Xi = 1,2 \}$:

$$\left| \varphi^{J_0 M_0}(t) \right\rangle = \sum_{J\varepsilon} \begin{pmatrix} C_{F_1}^{J_0 M_0}(J\varepsilon, t) & 0 \\ 0 & C_{F_2}^{J_0 M_0}(J\varepsilon, t) \end{pmatrix} \begin{pmatrix} \left| \Xi = 1 J M_0 \varepsilon \right\rangle \\ \left| \Xi = 2 J M_0 \varepsilon \right\rangle \end{pmatrix} . \tag{9}$$

At later times the density matrix is then constructed by

$$\hat{\rho}(t) = \sum_{J_0 M_0} w_{J_0} \left|\varphi^{J_0 M_0}(t)\right\rangle\left\langle\varphi^{J_0 M_0}(t)\right| \\ = \sum_{J_0 M_0} w_{J_0} \sum_{J\varepsilon\Xi, J'\varepsilon'\Xi'} C_{F_\Xi}^{J_0 M_0}(J\varepsilon,t) C_{F_{\Xi'}}^{J_0 M_0 *}(J'\varepsilon',t) \left|\Xi J M_0 \varepsilon\right\rangle\left\langle\Xi' J' M_0 \varepsilon'\right| . \tag{10}$$

Here the structure of this density matrix should be noted. The interaction with the laser pulse generally introduces transitions between the electronic states, indicated by the presence of both quantum numbers $\Xi$ and $\Xi'$ in the expansion. By tracing the density matrix over the purely rotational degrees of freedom, one therefore can obtain a reduced electronic density matrix $\hat{\rho}^{el}(t)$, that describes the occupation and coherence between spin-orbit states $F_1$ and $F_2$

$$\hat{\rho}^{el}(t) = \sum_{J\varepsilon\Xi, J'\varepsilon'\Xi'} \left\langle \Xi' J' M_0 \varepsilon' \right| \hat{\rho}(t) \left| \Xi J M_0 \varepsilon \right\rangle . \tag{11}$$

The matrix representation of this reduced electronic density matrix with respect to the quantum number $\Xi$ explicitly reads

$$\hat{\rho}_{\Xi\Xi'}^{el}(t) = \sum_{J_0 M_0} w_{J_0} \sum_{J\varepsilon, J'\varepsilon'} C_{F_\Xi}^{J_0 M_0}(J\varepsilon,t) C_{F_{\Xi'}}^{J_0 M_0 *}(J'\varepsilon',t) . \tag{12}$$

We define the degree of coherence between the two electronic states with the reduced electronic density matrix as

$$\mathbb{C}(t) = \frac{|\hat{\rho}_{12}^{el}(t)|}{\sqrt{\hat{\rho}_{11}^{el}(t)\hat{\rho}_{22}^{el}(t)}} . \tag{13}$$

On the other hand, tracing over the electronic quantum number $\Xi$ will lead to the pure rotational density matrix of the ensemble. An important object in our analysis is

$$\hat{\rho}(\varphi,\theta,\chi,t) := \left\langle \varphi,\theta,\chi \,|\, \hat{\rho}(t) \,|\, \varphi,\theta,\chi \right\rangle , \tag{14}$$

which can be interpreted as the angular distribution of the electronic density matrix. To analyze the angular dependence of this object, we will express $\hat{\rho}(\varphi,\theta,\chi,t)$ in terms of multipoles. The Wigner *D* functions fulfill the following product rule, resulting from the group-addition theorem of the rotation group

$$\left\langle \varphi,\theta,\chi \,|\, J_1,\Omega_1,M_0 \right\rangle \left\langle J_2,\Omega_2,M_0 \,|\, \varphi,\theta,\chi \right\rangle \\ = \frac{\sqrt{2J_1+1}\sqrt{2J_2+1}}{8\pi^2} D_{M_0,\Omega_1}^{J_1}(\varphi,\theta,\chi) D_{M_0,\Omega_2}^{J_2}(\varphi,\theta,\chi)^* \\ = (-1)^{M_0-\Omega_2} \frac{\sqrt{2J_1+1}\sqrt{2J_2+1}}{8\pi^2} D_{M_0,\Omega_1}^{J_1}(\varphi,\theta,\chi) D_{-M_0,-\Omega_2}^{J_2}(\varphi,\theta,\chi) \\ = (-1)^{M_0-\Omega_2} \frac{\sqrt{2J_1+1}\sqrt{2J_2+1}}{8\pi^2} \sum_K C_{K,0}^{J_1,M_0;J_2,-M_0} C_{K,Q}^{J_1,\Omega_1;J_2,-\Omega_2} D_{0,Q}^K(\varphi,\theta,\chi) \delta_{Q,\Omega_1-\Omega_2} \tag{15}$$

where $C_{J,M}^{J_1,M_1;J_2,M_2}$ indicates the Clebsch-Gordan coefficients. We therefore get the following irreducible representation of the angular distribution of the electronic density matrix:

$$
\begin{aligned}
\hat{\rho}(\varphi,\theta,\chi,t) &= \langle \varphi,\theta,\chi \,|\, \hat{\rho}(t) \,|\, \varphi,\theta,\chi \rangle \\
&= \sum_{J_0 M_0} w_{J_0} \sum_{J\varepsilon\Xi, J'\varepsilon'\Xi'} C_{F_\Lambda}^{J_0M_0}(J\varepsilon,t) C_{F_{\Lambda'}}^{J_0M_0 *}(J'\varepsilon',t) \langle \varphi,\theta,\chi \,|\, \Xi J M_0 \varepsilon \rangle \langle \Xi' J' M_0 \varepsilon' \,|\, \varphi,\theta,\chi \rangle \\
&= \sum_{\Xi\Xi'} \sum_{J_0 M_0} w_{J_0} \sum_{KQ=-3,3} [\, f_{K,Q}^{\Xi\Xi'}(J_0 M_0,t) D_{0,Q}^{K}(\varphi,\theta,\chi) + c.c.] \qquad (16) \\
&= \sum_{\Xi\Xi'} \sum_{KQ=-3,3} f_{K,Q}^{\Xi\Xi'}(t) D_{0,Q}^{K}(\varphi,\theta,\chi) + c.c. \\
&=: \sum_{\Xi\Xi'} \hat{\rho}_{\Xi\Xi'}(\varphi,\theta,\chi,t)
\end{aligned}
$$

Here we give an expansion of the occupations of the angular degrees of freedom in terms of matrix elements with respect to the quantum numbers $\Xi$ and $\Xi'$, that characterize the electronic state. The object $\hat{\rho}_{\Xi\Xi'}(\varphi,\theta,\chi,t)$ can therefore be interpreted as the angular distribution of the reduced electronic density matrix. The above expression has the structure of the expansion of the density matrix in terms of irreducible tensor operators, only that we limited the expression to the diagonal terms of the rotational degrees of freedom. The coefficients in such an expansion are then usually referred to the state multipoles of the density matrix [48]. Here, the coefficients $f_{K,Q}^{\Xi\Xi'}(t)$ can therefore be interpreted as the state multipoles of the “electronic” density matrix. $f_{K,Q}^{11}(t)$ and $f_{K,Q}^{22}(t)$ therefore physically refer to the angular multipoles-expansion coefficients of the occupations of electronic states $F_1$ and $F_2$, respectively. $f_{K,Q}^{12}(t)$ and $f_{K,Q}^{21}(t)$ describe the multipoles of the electronic coherence between states $F_1$ and $F_2$. The values for those generalized multipoles will be studied in the next section. Note that under the experimental conditions the system is not highly excited, *i.e.*, $a_J \sim 1$, $b_J \sim 0$. In this case $f_{K,Q}^{11}(t)$ and $f_{K,Q}^{22}(t)$ are dominated by the terms with $Q$=0, while $f_{K,Q}^{12}(t)$ is dominated by the terms with $Q = \pm 2$. In the present case, the contribution of $K$ is restricted to even numbers, since only alignment, that means no orientation, is achieved by the interaction. Since $D_{0,Q=0}^{K=2}(\varphi,\theta,\chi) = \frac{1}{2}(3\cos^2\theta - 1)$, $f_{K=2,Q=0}^{11}(t)$ and $f_{K=2,Q=0}^{22}(t)$ are linearly related to the time-dependent expectation value $\langle \cos^2\theta \rangle$ in the subspace of $F_1$ and $F_2$ states, respectively. As Eq. (16) also clearly shows, the quantum number $M_0$ is conserved under the interaction with the linearly polarized field. This means that $\hat{\rho}(\varphi,\theta,\chi,t)$ does not have a specific dependence on the angle $\varphi$. However, since $Q$ is in general not equal to zero, $\hat{\rho}(\varphi,\theta,\chi,t)$ depends not only on the angle $\theta$, but also on the angle $\chi$.

In order to study the effect of the HHG probe pulse under different direction of the polarization, a rotation of the density matrix becomes necessary. Let's define $\mathbf{R}(\alpha,\beta,\gamma)$ the rotational operator inducing finite rotations by the Euler angles $(\alpha,\beta,\gamma)$ with respect to the laboratory frame; the rotated angular distributions of the density matrix $\hat{\rho}^{\mathbf{R}(\alpha,\beta,\gamma)}(\varphi,\theta,\chi,t)$ can then be obtained by

$$\begin{aligned}
&\hat{\rho}^{\mathbf{R}(\alpha,\beta,\gamma)}(\varphi,\theta,\chi,t) \\
&=\left\langle \varphi,\theta,\chi \,|\, \mathbf{R}(\alpha,\beta,\gamma)\hat{\rho}(t)\mathbf{R}(\alpha,\beta,\gamma) \,|\, \varphi,\theta,\chi \right\rangle \\
&=\sum_{\Xi\Xi'} \hat{\rho}_{\Xi\Xi'}^{\mathbf{R}(\alpha,\beta,\gamma)}(\varphi,\theta,\chi,t) \qquad , \\
&=\sum_{\Xi\Xi'} \sum_{KQ=-3,3} f_{K,Q}^{\Xi\Xi'}(t)\mathbf{R}(\alpha,\beta,\gamma)D_{0,Q}^{K}(\varphi,\theta,\chi)+c.c.
\end{aligned} \tag{17}$$

and

$$\mathbf{R}(\alpha,\beta,\gamma)D_{0,Q}^{K}(\varphi,\theta,\chi)=\sum_{\mathrm{X}} D_{\mathrm{X},Q}^{K}(\varphi,\theta,\chi)D_{\mathrm{X},0}^{K}(\alpha,\beta,\gamma)\,. \tag{18}$$

**II.C The probe process: high-harmonic generation**

In this subsection we describe the HHG of the delayed probe pulse $\boldsymbol{\varepsilon}_2(t-\tau)$ in the prepared molecular ensemble. The HHG process is studied as a function of the delay time τ between the pump and probe pulses, measured with respect to the peaks of the field envelope. Since relatively slow wavepacket dynamics is probed by the HHG pulse – the rotational wavepacket has a revival time of about 20 ps and the relatively slow spin-orbit period is about 280 fs – the density matrix of Eq. (10), describing the rotational/spin-orbit wavepacket, will be considered as frozen during the interaction with probe pulse, that has a duration of about 30 fs. For every time delay τ, the density matrix $\hat{\rho}(\tau)$ therefore defines an initial state for the HHG probe process. In order to study the HHG process, the rigid-rotor density matrix of Eq. (10) has to be extended to also include the electronic degrees of freedom

$$\hat{\rho}^{total}(\tau;t)=\sum_{J_0M_0} w_{J_0} \left|\Psi^{J_0M_0}(\tau;t)\right\rangle\left\langle\Psi^{J_0M_0}(\tau;t)\right|. \tag{19}$$

Here the wave function $\left|\Psi^{J_0M_0}(\tau;t)\right\rangle$ denotes the extension of the rotational wave function $\left|\varphi^{J_0M_0}(t)\right\rangle$ by the electronic degrees of freedom. The rotational part of the wavefunction is assumed to be constant during the HHG process, τ is hence a parameter which defines the rotational /spin-orbit initial state for the HHG probe process. The electronic degrees of freedom are acting under the influence of the probe pulse. To determine the temporal evolution of the density matrix under the action of the HHG probe pulse, within the single-active-electron approximation, we have to solve for the propagation of the "initial" states $\left|\Psi^{J_0M_0}(\tau;t)\right\rangle$ under the evolution of the electronic Hamiltonian

$$i\frac{\partial}{\partial t}\left|\Psi(\tau;t)\right\rangle = [H_e - \boldsymbol{\mu}\cdot\boldsymbol{\varepsilon}_2(t)]\left|\Psi(\tau;t)\right\rangle, \tag{20}$$

where $H_e$ is the field-free Hamiltonian in the molecular frame, including both electronic and rotational parts, $\boldsymbol{\mu}$ is the electronic dipole operator. The probe field $\boldsymbol{\varepsilon}_2(t)$ has a pulse duration of ~30 fs. The polarization axis of the probe pulse is arbitrary and will be varied. In particular we will study parallel and orthogonal polarization directions of the pump $\boldsymbol{\varepsilon}_1(t)$ and the probe field $\boldsymbol{\varepsilon}_2(t)$.

The typical values for the probe peak intensities are in the range of 1.0-1.5×$10^{14}$ W/cm$^2$ [39]. At these intensities, the bound-state is gradually depleted due to the strong-field ionization [30], but the main effect of depletion is to reduce the overall intensity of high-harmonic emission [34]. Since we are not interested in the absolute intensities but only in the time-dependent relative intensities we assume that the electronic bound-state wave function $\left|\Psi^{J_0M_0}(\tau;t)\right\rangle$ is not depleted. Moreover, the probe field does not directly couple to the rotational degrees of freedom [30]. Under these conditions, we expand the rotational-electronic wave function in

$$\begin{aligned}
&\left|\Psi^{J_0M_0}(\tau;t)\right\rangle \\
&= \sum_{J\varepsilon} e^{i(I_p+E_J^{(1)})t} C_{F_1}^{J_0M_0}(J\varepsilon,\tau)\left|F_1\,;JM_0\varepsilon\right\rangle + \sum_{J\varepsilon} e^{i(I_p+E_J^{(2)}+\Delta E)t} C_{F_2}^{J_0M_0}(J\varepsilon,\tau)\left|F_2\,;JM_0\varepsilon\right\rangle, \\
&+ e^{iI_pt}\sum_{J_cM_c}\int d^3\mathbf{k}\, C_c(\vec{k},J_cM_c,t)\left|\mathbf{k};J_cM_c\right\rangle
\end{aligned} \tag{21}$$

where $I_p$=9.26 eV is the ground-state ionization potential, $\Delta E$=0.015 eV is the energy difference between the two lowest rotational state of each of the electronic states. Since the rotational energies $E_J^{(\Xi)}$ and the energy difference $\Delta E$ between these two states are much smaller than the ionization potential $I_p$ ( $E_J^{(\Xi)}$ ~$10^{-5}$ $I_p$, $\Delta E$~$10^{-4}$ $I_p$), it is a good approximation to neglect both $E_J^{(\Xi)}$ and $\Delta E$ in Eq. (21). The coefficients $C_{F_1}^{J_0M_0}$ and $C_{F_2}^{J_0M_0}$ are determined by the interaction with the pump (alignment) pulse and we assume that these coefficients are not modified during the probe pulse interaction; $\left\{\left|F_\Xi\,;JM\varepsilon\right\rangle\right\}$ $(\Xi = 1,2)$ denote the combined electronic-rotational eigenstates, *i.e.*, they include the electronic radial wavefunction, in contrast to $\{|\Xi JM\varepsilon\rangle\}$ $(\Xi = 1,2)$ of Eq. (2), which only contains the rotational degrees of freedom. The superscript $J_0M_0$ indicates the "initial" electronic-rotational wavefunction $|F_1\ ; J_0M_0\varepsilon\rangle$; $|\mathbf{k}; J_cM_c\rangle = |\mathbf{k}\rangle \otimes |J_cM_c\rangle$, where $|\mathbf{k}\rangle$ denotes the electronic continuum states of asymptotic wave vector $\mathbf{k}$, $|J_cM_c\rangle$ denotes the remaining rotational states of the ionic core. The continuum coefficients $C_c(\mathbf{k}, J_cM_c, t)$ are calculated within the strong-field approximation [34] and are approximated by

$$C_c(\mathbf{k}, J_cM_c, t) = i\int_0^t dt' \sum_{J\varepsilon\Xi} C_{F_\Xi}^{J_0M_0}(J\varepsilon,\tau)\left\langle \mathbf{k}'; J_cM_c \,|\, \boldsymbol{\mu}\cdot\hat{\boldsymbol{\varepsilon}}_2(t') \,|\, F_\Xi\,; JM_0\varepsilon\right\rangle e^{-iS(t,t')}\,. \tag{22}$$

Here $S(t,t') = \int_{t'}^{t} dt''[(\mathbf{k}+\mathbf{A}(t)-\mathbf{A}(t''))^2/2 + I_p]$ denotes the part of the classical action of the continuum electron acquired during the interaction with the probe laser pulse between the moment of ionization at time *t*' and recollision at time *t*, $\mathbf{k}' = \mathbf{k} + \mathbf{A}(t) - \mathbf{A}(t')$ is the electronic momentum at time of the recollision and $\mathbf{A}(t) = -\int_0^t \boldsymbol{\varepsilon}(t')dt'$ denotes the vector potential of the electric field. As any approximation to the full time-dependent many-body Schrödinger equation, the strong-field approximation is not gauge invariant. Both length gauge and velocity gauge have their respective merits in different situations, and the length gauge is suggested as a preferred gauge in describing molecular orientation effects on the HHG [49]. We therefore employed the length gauge in this work. Moreover the use of the conventional strong-field approach neglects the interaction of the continuum electron with the short-range and Coulomb potentials of the molecular cation. We have therefore replaced the traditional plane-wave-based recombination matrix elements of the SFA with matrix elements calculated within a proper description of the electron-cation scattering process (see Eq. (34) below and Ref. [50]).

The emitted HHG spectrum of linear polarization along the direction $\hat{n}$ in the laboratory frame is then determined by the expectation value of the electronic dipole $\boldsymbol{\mu}$ along the direction $\hat{n}$. The dipole operator $\boldsymbol{\mu}$ here refers to the molecular frame, whereas $\boldsymbol{n}$ is measured within the laboratory frame. Explicitly, the dipole expectation value at time *t*, for a probe pulse of time delay $\tau$ is given by

$$
\begin{aligned}
& d(t,\tau) \\
& = tr[\hat{\rho}^{total}(t)\boldsymbol{\mu}\cdot\hat{n}] \\
& = \sum_{J_0M_0} w_{J_0} \left\langle \Psi^{J_0M_0}(\tau;t) \middle| \boldsymbol{\mu}\cdot\hat{n} \middle| \Psi^{J_0M_0}(\tau;t) \right\rangle \\
& = \sum_{J_0M_0} w_{J_0} \sum_{J\varepsilon, J_cM_c\Xi} C_{F_\Xi}^{J_0M_0\,*}(J\varepsilon,\tau) \int d^3\mathbf{k} C_c(\mathbf{k}, J_cM_c, t) \left\langle F_\Xi; JM_0\varepsilon \,|\, \boldsymbol{\mu}\cdot\hat{n} \,|\, \mathbf{k}; J_cM_c \right\rangle + c.c.
\end{aligned} \quad . \tag{23}
$$

The next step in the derivation of the final expression is to insert the expansion of the continuum coefficients, and using the closure relation [30]

$$
\sum_{J_cM_c} |J_cM_c\rangle\langle J_cM_c| = 1 = \int d\hat{R} \left|\hat{R}\right\rangle\left\langle\hat{R}\right|, \tag{24}
$$

where $\left\{\hat{R}\right\}$ denote the Euler angles $(\varphi,\theta,\chi)$ defined with respect to the lab frame. We define $\left|\mathbf{k};\hat{R}\right\rangle = |\mathbf{k}\rangle \otimes \left|\hat{R}\right\rangle$. The induced dipole is then given by

$$
\begin{aligned}
& d(t,\tau) \\
& = i \sum_{J_0M_0} w_{J_0} \int d\hat{R} \sum_{J\varepsilon\Xi\Xi'} C_{F_\Xi}^{J_0M_0\,*}(J\varepsilon,\tau) \sum_{J'\varepsilon'} C_{F_{\Xi'}}^{J_0M_0}(J'\varepsilon',\tau) \int d^3\mathbf{k} \left\langle F_\Xi; JM_0\varepsilon \middle| \boldsymbol{\mu}\cdot\hat{n} \middle| \mathbf{k};\hat{R} \right\rangle \cdot \\
& \int_0^t dt' \left\langle \mathbf{k}';\hat{R} \middle| \boldsymbol{\mu}\cdot\hat{\boldsymbol{\varepsilon}}_2(t') \middle| F_{\Xi'}; J'M_0\varepsilon' \right\rangle e^{-iS(t,t')} + c.c.
\end{aligned} \tag{25}
$$

As discussed earlier, Hund's case (a) is an appropriate description for the system*, the total molecular wavefunctions can therefore be approximated by products of electronic spin-orbit states ($\left|F_\Xi\right\rangle$, includes the angular and spin degrees of freedom) and the rotational rigid rotor wavefunctions as

$$
\left|F_\Xi; JM_0\varepsilon\right\rangle = \left|F_\Xi\right\rangle \otimes |\Xi JM_0\varepsilon\rangle \,. \tag{26}
$$

Where the electronic wavefunctions are given by [8]

$$
\begin{cases}
|F_1\rangle = \dfrac{1}{\sqrt{2}}(\left|\pi^+\beta\right\rangle + \left|\pi^-\alpha\right\rangle) \\
|F_2\rangle = \dfrac{1}{\sqrt{2}}(\left|\pi^+\alpha\right\rangle + \left|\pi^-\beta\right\rangle)
\end{cases} . \tag{27}
$$

Here $\left|\pi^\pm\right\rangle = |\pi_x\rangle \pm i|\pi_y\rangle$, and $\pm$ in the subscript stand for the orbital angular momentum projection $|\Lambda = \pm 1\rangle$. $|\alpha\rangle / |\beta\rangle$ stands for $|\Sigma = \pm 1/2\rangle$, respectively. $\pi_x$ and $\pi_y$ are two degenerate components of the singly-occupied $\pi$ molecular orbital of NO [8]. Note that the dependence of $\pi^+$ and $\pi^-$ on $\chi$ have the form $e^{-i\chi}$ and $e^{i\chi}$, respectively [43]. Having

* Our theory can be extended to the other Hund's cases. For this purpose, it is convenient to introduce a new coupled angular momentum basis according to the cases (b), (c), (d), or (e), that can be achieved by a unitary transformation of the coupled angular momentum basis of case (a) [51] H. Lefebvre-Brion, and R. W. Field, *The Spectra and Dynamics of Diatomic Molecules: Revised and Enlarged Edition* (Academic Press, 2004). .

introduced the separation of the rotational from the electronic degrees of freedom, the expression for the dipole expectation value can be simplified and is given as an integral of a reduced rotational/electronic density matrix and an pure electronic dipole expectation value:

$$d(t,\tau) \simeq \int d\hat{R} \sum_{\Xi\Xi'} \rho_{\Xi\Xi'}(\hat{R},\tau) D_{\Xi\Xi'}(\hat{R},t) . \quad (28)$$

Here, we defined the reduced electronic density matrix

$$\rho_{\Xi\Xi'}(\hat{R},\tau) = \sum_{J_0 M_0} w_{J_0} \sum_{J\varepsilon} C_{F_\Xi}^{J_0 M_0}(J\varepsilon,\tau) \left\langle \hat{R} \,|\, \Xi J M_0 \varepsilon \right\rangle [\sum_{J'\varepsilon'} C_{F_{\Xi'}}^{J_0 M_0}(J'\varepsilon',\tau) \left\langle \hat{R} \,|\, \Xi' J' M_0 \varepsilon' \right\rangle]^* , \quad (29)$$

that is equivalent to the expression in Eq. (16). The reduced electronic matrix elements $D_{\Xi\Xi'}$ of Eq. (28) are containing electronic dipole transition matrix elements for tunnel ionization from state $F_{\Xi'}$ and the recombination to state $F_\Xi$

$$D_{\Xi\Xi'}(\hat{R},t) = i \int d^3\mathbf{k} \left\langle F_\Xi \,|\, \boldsymbol{\mu}\cdot\hat{n} \,|\, \mathbf{k} \right\rangle \int_0^t dt' \left\langle \mathbf{k}' \,|\, \boldsymbol{\mu}\cdot\hat{\boldsymbol{\varepsilon}}_2(t') \,|\, F_{\Xi'} \right\rangle e^{-iS(t,t')} + c.c. . \quad (30)$$

This reduced matrix element depends on the Euler angles, since the electronic dipole operator is defined with respect to the molecular frame, whereas the polarization directions of the applied field and the emitted light are defined with respect to the laboratory frame. Moreover, as will be shown in the next paragraph and appendix A, the matrix elements between the states $F_1$ and $F_2$ introduce a $\chi-$dependent phase factor. In the expression of Eq. (28), one immediately sees that the HHG signal has contributions from four different electronic channels, depicted in Fig.1a; Contributions to the dipole expectation value containing $\rho_{\Xi\Xi}(\hat{R},\tau)$ $(\Xi = 1,2)$ correspond to the single-channel HHG process (direct channels). In addition to these conventional HHG channels, the dipole expectation value also depends on a coherent cross channel contribution, determined by the electronic coherences $\rho_{12}(\hat{R},\tau)$ and $\rho_{21}(\hat{R},\tau)$ between the $F_1$ and $F_2$ states.

The HHG spectrum at delay time τ is determined by the Fourier transform of $d(t,\tau)$ with respect to $t$

$$d(\omega,\tau) = \int d\hat{R} \sum_{\Xi\Xi'} \rho_{\Xi\Xi'}(\hat{R},\tau) D_{\Xi\Xi'}(\hat{R},\omega) , \quad (31)$$

where $D_{\Xi\Xi'}(\hat{R},\omega)$ denotes the Fourier transform of the product of the tunnel ionization and recombination matrix elements of Eq. (30). Eq. (31) describes the situation of parallel polarizations of the pump and probe laser. For the general case of relative angles of the two polarization directions given by the Euler angles $(\alpha,\beta,\gamma)$, the dipole expectation value is given

by

$$d(\omega,\tau)=\int d\hat{R}\sum_{\Xi\Xi'}\hat{\rho}_{\Xi\Xi'}^{\mathbf{R}(\alpha,\beta,\gamma)}(\hat{R},\tau)D_{\Xi\Xi'}(\hat{R},\omega)\,. \tag{32}$$

It now remains to give an explicit expression for $D_{\Xi\Xi'}(\hat{R},\omega)$. As presented in the appendix A, if both fine structure components $\pi^{+}$ and $\pi^{-}$ are assumed to have the same radial wave function, the matrix $D_{\Xi\Xi'}(\hat{R},\omega)$ is given by

$$\begin{aligned}D_{\Xi\Xi'}(\hat{R},\omega)&\simeq i\int d^3\mathbf{k}\left\langle\Phi_{\mathrm{HOMO}}(\theta,\chi=0)\,|\,\boldsymbol{\mu}\cdot\hat{n}\,|\,\mathbf{k}\right\rangle\\&\times\int_0^t dt'\left\langle\mathbf{k}'\,|\,\boldsymbol{\mu}\cdot\boldsymbol{\varepsilon}_2(t')\,|\,\Phi_{\mathrm{HOMO}}(\theta,\chi=0)\right\rangle e^{-iS(t,t')}\frac{e^{-i2(\Xi-\Xi')\chi}+e^{i2(\Xi-\Xi')\chi}}{2}+c.c.\end{aligned}\,. \tag{33}$$

Following the concepts introduced in molecular-orbital tomography [52, 53] or the quantitative rescattering method (QRS) [37], $D(\hat{R},\omega)$ can be expressed by the following product [54],

$$D(\hat{R},\omega)=\sqrt{\Gamma(\hat{R})}a_{ewp}(\omega)d_{rec}(\hat{R},\omega)\,, \tag{34}$$

where $\Gamma(\hat{R})$ is the angle-dependent strong field ionization rate, $a_{ewp}(\omega)$ is called the complex photoelectron wavepacket, $\Gamma(\hat{R})\left|a_{ewp}(\omega)\right|^2$ describes the flux of the returning electrons [37], and $d_{rec}(\hat{R},\omega)$ is the complex dipole recombination matrix element [50]. So Eq. (33) can then be rewritten as

$$D_{\Xi\Xi'}(\varphi,\theta,\chi,\omega)=\sqrt{\Gamma(\theta,\chi=0)}a_{ewp}(\omega)d_{rec}(\theta,\chi=0,\omega)\frac{e^{-i2(\Xi-\Xi')\chi}+e^{i2(\Xi-\Xi')\chi}}{2}\,. \tag{35}$$

Note that appendix B provides an alternative derivation of Eq. (35). The expectation value of the dipole moment of Eq. (32) for the general case of pump and probe polarization can be explicitly expressed in terms of the irreducible tensor components of the density matrix and reads

$$\begin{aligned}d(\omega,\tau)&=4\pi^2\int d\theta\sin\theta\sqrt{\Gamma(\theta,\chi=0)}a_{ewp}(\omega)d_{rec}(\theta,\chi=0,\omega)\\&\times\sum_K d_{0,0}^K(\beta)\mathrm{Real}\{[f_{K,-2}^{11}(\tau)+f_{K,-2}^{22}(\tau)+f_{K,-2}^{12}(\tau)]d_{0,-2}^K(\theta)\\&+2[f_{K,0}^{11}(\tau)+f_{K,0}^{22}(\tau)+f_{K,0}^{12}(\tau)]d_{0,0}^K(\theta)+[f_{K,2}^{11}(\tau)+f_{K,2}^{22}(\tau)+f_{K,2}^{12}(\tau)]d_{0,2}^K(\theta)\}\end{aligned}\,. \tag{36}$$

*Note that the integrations of $\varphi$ and $\chi$ in Eq. (32) only select the term with $\mathrm{X}=0$ in

* It should be noted that in the theoretical treatment of Ref. [39] D. Baykusheva, P. Kraus, S. B. Zhang, N. Rohringer, and H. J. Worner, Faraday Discuss. **171**, 113 (2014)., $\chi$ was set to be zero in the density matrix. Compared to the exact expression of Eq. (36), the evaluation of the dipole expression value by Eq.(28) of Ref. [39] ibid. resulted in additional contributions from terms proportional to $\mathrm{Real}(f_{K,Q}^{\Xi\Xi'})$ with $Q=\pm1,\pm3$ in $\hat{\rho}^{\mathbf{R}(\alpha,\beta,\gamma)}$, those terms are much smaller than the terms proportional to $\mathrm{Real}(f_{K,Q}^{\Xi\Xi'})$ with $Q=0,\pm2$ in $\hat{\rho}^{\mathbf{R}(\alpha,\beta,\gamma)}$. Therefore, the numerical differences of the HHG spectra between the exact treatment and the approximated treatment by neglecting the dependence on the angle $\chi$ of Ref. [39] ibid.turn out to be negligible, giving virtually the same

$D^K_{X,0}(\alpha,\beta,\gamma)$ and the terms with $Q=0,\pm 2$ in $\hat{\rho}^{\mathbf{R}(\alpha,\beta,\gamma)}$ and $D^K_{0,0}(\alpha,\beta,\gamma)=d^K_{0,0}(\beta)$. Our numerical evaluation shows that $\mathrm{Real}(f^{11}_{K,Q=\pm2}(t))$, $\mathrm{Real}(f^{22}_{K,Q=\pm2}(t))$ and $\mathrm{Real}(f^{12}_{K,Q=0}(t))$ are about three orders of magnitude smaller than $\mathrm{Real}(f^{11}_{K,Q=0}(t))$, $\mathrm{Real}(f^{22}_{K,Q=0}(t))$ and $\mathrm{Real}(f^{12}_{K,Q=\pm2}(t))$, respectively. Taking only into account the main contributions of $\mathrm{Real}(f^{11}_{K,Q}(t))$, $\mathrm{Real}(f^{22}_{K,Q}(t))$ and $\mathrm{Real}(f^{12}_{K,Q}(t))$ with the irreducible components of $Q=0$ and $Q=\pm2$, respectively, Eq.(36) can be approximated by

$$d(\omega,\tau)\simeq 4\pi^2\int d\theta\sin\theta\sqrt{\Gamma(\theta,\chi=0)}a_{ewp}(\omega)d_{rec}(\theta,\chi=0,\omega)[W^{MC}(\theta,\beta,\tau)+W^{CC}(\theta,\beta,\tau)], \qquad (37)$$

where we have defined the purely geometric quantities giving rise to the direct HHG channel

$$W^{DC}(\theta,\beta,\tau)=\sum_K 2d^K_{0,0}(\beta)\mathrm{Real}[f^{11}_{K,0}(\tau)+f^{22}_{K,0}(\tau)]d^K_{0,0}(\theta), \qquad (38)$$

And the component giving rise to the cross channel HHG

$$W^{CC}(\theta,\beta,\tau)=\sum_K d^K_{0,0}(\beta)\mathrm{Real}[f^{12}_{K,-2}(\tau)d^K_{0,-2}(\theta)+f^{12}_{K,2}(\tau)d^K_{0,2}(\theta)]. \qquad (39)$$

Eq. (36) and Eq. (37) show that the induced dipole only depends on the angle $\beta$, defined as the relative angle between the polarizations of the pump and probe pulses. Note that in the case of parallel polarization $d^K_{0,0}(\beta=0)$ is 1 for all *K*; and in the case of perpendicular polarizations $d^K_{0,0}(\beta=\pi/2)$ is 1, -1/2, 3/8 and -5/16 for *K* = 0, 2, 4 and 6, respectively. Eq. (36) and Eq. (37) also show that the intensities of different harmonics are mainly dominated by the recombination dipole-matrix elements, which depend on energy of the recombining continuum electron, and hence on the harmonic order. In Ref. [39] we showed both experimental and theoretical results of the entire harmonic plateau. Here, we focus on the delay dependence of the harmonic yield of two typical harmonics of the plateau region. The harmonics are chosen in a demonstrative way, to highlight the strong dependence on electronic degrees of freedom in 15$^{th}$ harmonic and the strong sensitivity of rotational dynamics in 9$^{th}$ harmonic.

## III. Discussion of numerical results

Our calculations are performed with parameters according to the experimental conditions.

numerical results.

Within the range of experimental conditions, we choose parameters to best fit the experimental data: the molecular initial rotational temperature was set to be 10 K; pump pulse parameters: wavelength 800 nm, pulse duration 60 fs, and peak intensity $4\times10^{13}$ W/cm$^2$.

The $J$ state populations before and after the pump pulse of $F_1$ and $F_2$ electronic states are shown in Fig. 2. Initially, only $F_1$ is occupied and the populations of different $J$ states satisfy the Boltzmann distribution for an initial temperature $T$=10 K. The rotational states with $J$=1/2 and 3/2 possess the largest weight of about 39%, and only states up to $J_{max}$ = 9/2 are considerably occupied; after the interaction with the pump pulse, the system is rotationally excited to higher rotational states. States up to $J_{max}$ = 17/2 are considerably occupied for $F_1$, and the state with $J$=5/2 possesses the largest weight of about 23%. The rotational states of $F_2$ are weakly excited. The inset of Fig.2 shows the time-dependent total occupation probabilities of the $F_1$ and $F_2$ states, *i.e.*, $\hat{\rho}_{11}^{el}(t)$ and $\hat{\rho}_{22}^{el}(t)$ (see Eq. (12)). The total excitation fraction from the electronic states $F_1$ to $F_2$ is about 4% for molecules exposed to the peak intensity of the laser pulse. Note that although focal-volume averaging will reduce the averaged excitation fraction, 4% is considerably larger than the number of 0.2% reported in Ref. [8]. In Ref. [8] we only treated the first term of the interaction Hamiltonian of Eq. (7), which resulted in an incomplete treatment of the pump interaction. The relative importance of the two main terms in the interaction Hamiltonian of Eq. (7) is discussed in the following. Ref. [39] correctly included all terms of the interaction Hamiltonian and focal-volume averaging.

To get an idea about the angular distribution of the occupation probabilities of the reduced electronic density matrix as a function of time, we analyze the state multipoles $\mathrm{Real}(f_{K,Q}^{\Xi\Xi}(t))$ (see Eq. (16)). Fig. 3 shows the temporal evolution of $\mathrm{Real}(f_{K,0}^{11}(t))$ and $\mathrm{Real}(f_{K,0}^{22}(t))$ for $K$= 0, 2, 4 and 6. The absolute values of those state multipoles drop fast with increasing $K$. The isotropic part ($K$=0) is constant before and after the interactions with the pump pulse. The state multipoles for higher $K$ show complex dynamics. For $K$=2, $\mathrm{Real}(f_{K,0}^{11}(t))$ and $\mathrm{Real}(f_{K,0}^{22}(t))$ are linearly related to the expectation value of $\cos^2\theta$ in the subspace of $F_1$ and $F_2$ states, *i.e.*, the typical measure for the degree of alignment of the molecular ensemble. Clearly seen is the rotational revival structure at around 20 ps. The state multipoles connected to the electronic coherence between states $F_1$ and $F_2$ show much faster dynamics. Fig. 4 shows the temporal variation of the state multipoles $\mathrm{Real}(f^{12}_{K,\pm 2}(t))$ for $K$= 2, 4 and 6. Clearly visible is the fast modulation, corresponding to a period of about 0.28 ps, which is the spin-orbit period corresponding to the energy separation of the $F_1$ and $F_2$ states of about 120 $cm^{-1}$.

In order to understand how these angular distributions of the electronic density matrix determine the HHG spectrum as a function of delay time τ, we have a closer look at the approximated expression for $d(\omega,\tau)$ given in Eq. (37). The HHG contributions resulting from the direct channel, for which ionization and recombination proceed from and to the same electronic state, are determined by the purely geometric quantity $W^{DC}(\theta,\beta,\tau)$ . In Fig. 5 we plot the quantity $W^{DC}(\theta,\beta,\tau)$ for parallel (β = 0) and perpendicular (β = π/2) polarization of the pump and the probe laser fields. For a delay time around τ = 5.2 ps we see maxima in the distribution $W^{DC}(\theta,\beta,\tau)$ around θ = 0$^{o}$ and 180$^{o}$. These peaks correspond to the main rotational revival structure of the wavepacket also visible in Fig. 3. The direct channel contribution to HHG therefore shows a maximum at the rotational revivals, when probed by a field of parallel polarization. Switching to the perpendicular polarization (β = π/2), molecules aligned at θ=90$^{o}$ give the highest signal contribution at τ = 5.2 ps. For β = π/2 the highest HHG yield is observed at delay times τ = 4.8 ps, coinciding with the anti-aligned ensemble in Fig. 3 (red curve, corresponding to the $K$ = 2 contributions). An interesting point is also to look at the relative modulation strength of the direct channel contribution at times 4.8 and 5.2 ps and compare the cases of parallel and perpendicular polarizations. By comparing the numbers, it is seen that for perpendicular polarizations, the relative modulation of the signal is decreased by a factor of about 2. This variation of the modulation strength comes from the dependence of $d^{K}_{0,0}(\beta)$ for different $K$.

Now, let's turn to the cross channel contributions of HHG, determined by the quantity $W^{CC}(\theta,\beta,\tau)$ of Eq. (39). This quantity, shown in Fig. 6 for β = 0 and β = π/2, is related to the electronic coherence between states $F_1$ and $F_2$, and shows fast oscillations, on a time scale of about 280 *f*s, *i.e.*, the spin-orbit period. A direct comparison between the case of parallel β = 0 or perpendicular polarization β = π/2 clearly shows a π phase shift. At the rotational wave-packet revival time at around 4.8 ps, $W^{CC}$ shows a clear maximum for β = 0 and a minimum for β = π/2. The angular distribution of this maximum and minimum is almost flat. Generally, the angular variation at a fixed time of $W^{CC}$ is much weaker, as compared to the angular variation of the direct channel contribution $W^{DC}$.

We now analyze the temporal dependence of different harmonics. Fig. 7 shows the HHG signal of the 9$^{th}$ harmonic (H9) for parallel polarization of pump and probe laser (β = 0) as a function of the delay time between the two pulses. Our calculations (upper panels) are compared to the experimental results (lower panels). Shown are time traces (left panels) and their Fourier transforms (FFT) (right panels). Our calculations capture most of the relevant dynamics and theory and experiment can be compared on a quantitative level. The agreement is remarkably good. The oscillations of the electronic coherence are clearly visible within the first few ps. With increasing time their modulation becomes weaker due to dephasing of the electronic wavepacket: the energy splitting of the $F_1$ and $F_2$ states, pertaining to a total angular momentum state *J*, depends weakly on *J*, which induces a dephasing for long delays. To underline the importance of the coherent cross channel HHG process, governed by off-diagonal parts of the density matrix, we also plot the delay-dependent H9 yield omitting this channel (black line). Although the rotational revival structure is reproduced in this case, omission of the cross channel coupling results in a loss of the modulation characteristic of the spin-orbit dynamics. The Fourier transform of the time traces gives a good way of comparing theory to experiment [8]. The peaks around 30 cm$^{-1}$ correspond to the rotational wavepacket coherence associated with a change of angular quantum number $\Delta J = 2$. Peaks around 120 cm$^{-1}$ result from the electronic wavepacket coherence [8]. The peaks in the range from ~60 cm$^{-1}$ to ~100 cm$^{-1}$ correspond to the rotational excitations by $\Delta J = 4$ [8, 39].

Similarly, Fig. 8 shows the delay-dependent HHG signal of the 15$^{th}$ harmonic (H15) for β = 0 for both theory (upper panels) and experiment (lower panels), along with their corresponding Fourier transforms. In contrast to H9, H15 has a weak dependence on the rotational wavepacket and is mainly dominated by the electronic coherence. This behavior is also captured by our theoretical results. The lower sensitivity to the rotational degrees of freedom for H15 can be explained by the relatively flat angular dependence of the recombination dipole-matrix elements for H15 [8].

We now study in more detail the dependence of the temporal HHG traces as a function of the probe-pulse polarization. Fig. 9 shows the HHG signals for H9 and H15 for parallel and perpendicular probe-pulse polarization. The experimental data are shown as well. The calculated signals reproduce the general structures of the experimental data. The most prominent difference is an observed π phase shift of the fast electronic modulations for β = 0 and π/2, that is recovered by the theory. This phase shift is directly related to the off-diagonal parts of the angular distributions of the density matrix shown in Fig. 6.

In the following we highlight the relative importance and influence of the two different terms in the interaction Hamiltonian of Eq. (7). We study the case where only terms proportional to $D^2_{0,0}$ in the interaction Hamiltonian are maintained, we refer to that as interaction term D0. In the second case, we study the interactions mediated by the term proportional to $D^2_{0,\pm2}$, interaction term D2. Fig. 10 shows the *J* state populations of the $F_1$ and $F_2$ states before and after the pump pulse for both interaction terms. The insets show the total occupations of the $F_1$ and $F_2$ sates as a function of time. Whereas the interaction term D0 only excites ~ 0.15 % of the population in the F1 state, D2 gives rise to an excitation fraction of ~4%. Hence, as already mentioned in the discussion of the matrix-representation of the interaction Hamiltonian, D2 is mainly responsible for transferring population to the spin-orbit excited states; while the rotational wavepackets are mainly induced by the interaction term D0. Therefore there is a clear separation of the interaction Hamiltonian: tensor components proportional to $D^2_{0,0}$ are responsible for rotational Raman excitations, whereas those proportional to $D^2_{0,\pm2}$ dominate electronic Raman scattering. Consistent conclusions can be drawn from the dynamic evolutions for the cross term of the reduced electronic density matrix $\hat{\rho}^{el}_{12}(t)$ and the defined coherence $\mathbb{C}(t)$ for interactions D0, D2 and the full interaction Hamiltonian of Eq. (7) in Fig. 11. It turns out that the modulation amplitude of $\hat{\rho}^{el}_{12}(t)$ for D0 and D0+D2 are similar and about ten times stronger than that for D2. The system only treating the interaction D0 shows a large electronic coherence (about 0.75 after the pump pulse), the inclusion of interaction D2 will significantly reduce the electronic coherence of the system (about 0.11 after the pump pulse), and the interaction term D2 alone results in a very small coherence (about 0.01 after the pump pulse). This can be explained by analyzing the interaction Hamiltonian Eq. (7) in more detail.

In the eigenbasis $\{|\Xi J M_0 \varepsilon\rangle, \Xi = 1,2\}$ expansion, the matrix elements related to the interaction terms $D^2_{0,0}$ or $D^2_{0,\pm 2}$ can be split into four main blocks: two direct blocks with transitions within the state $F_1$ and $F_2$, and two cross blocks, inducing transitions between $F_1$ and $F_2$. In each block, $D^2_{0,0}$ or $D^2_{0,\pm 2}$ is block diagonal for $\Delta J = 0,\ \pm 1,\ \pm 2$. For the interaction term proportional to $D^2_{0,0}$, the interaction is strong, i.e., non-perturbative, in the direct blocks, for which the matrix elements are proportional to $a_J a_{J'}$. This interaction in the direct blocks creates rotational wavepackets within one electronic subspace. Our initial state is in subspace $F_1$, $D^2_{0,0}$ mainly creates rotational wavepackets in that subspace. The interaction is weak, i.e., perturbative in the cross blocks, for which the matrix elements are proportional to $a_J b_{J'}$. This interaction in the cross blocks creates a small excitation from state $F_1$ to $F_2$. The non-perturbative part, hence creates a rotational wave-packet within $F_1$ which then in first order perturbation theory results in a small excitation fraction to electronic state $F_2$ by the transition matrix elements proportional to $a_J b_{J'}$, with well defined phases between states [55] of different angular momentum states *J* of manifold $F_1$ and *J'* of manifold $F_2$. Tracing over the rotational degrees of freedom to determine the reduced electronic density matrix hence yields a high degree of coherence between electronic states $F_1$ and $F_2$.

Let's now consider the interaction term proportional to $D^2_{0,\pm 2}$. For that interaction term, the dominant and non-perturbative contribution are the cross blocks of the interaction matrix, proportional to $a_J a_{J'}$; while the interaction matrix elements of the direct blocks are weak and proportional to $a_J b_{J'}$. The dominant interaction therefore induces a transition from state $F_1$ to $F_2$ along with excitation of a rotational wavepacket. Since the interaction is non-perturbative, there is no "fixed" phase relationship between rotational states of the lower and the upper electronic manifold. The electronic coherence of the excited electronic wavepacket is therefore smaller than that mediated by $D^2_{0,0}$.

An evidence for this can also be seen by comparing the occupations of the different rotational states before and after the pump-pulse interaction, that is shown in Fig. 10 for terms D0 and D2. In the case of D0 (upper panel of Fig. 10), one sees that the distribution in $F_2$ shiftes to states with higher $J$ with respect to the distribution of state $F_1$. The very small excitation fraction to $F_2$ results in a distribution of $J$ states after the pump-pulse interaction that is very similar to that of the lower state after the pump-pulse interaction. One can conclude that the rotational wavepackets pertaining to states $F_1$ and $F_2$ are similar, so that when tracing the density matrix over the total angular momentum $J$, a high degree of coherence is achieved. As seen in Fig. 11, the degree of coherence for interaction D0 right after the pump pulse is about 0.75. As a function of time, the degree of coherence decays. This is due to dephasing of the wavepacket, since the energy difference between electronic states $F_1$ and $F_2$ is dependent on the angular momentum number $J$ (see level system in Fig. 1). Interaction term D2 excites a larger excitation fraction to state $F_2$ (roughly 4%) along with excitation of a rotational wavepacket in manifold $F_2$, without really modifying the rotational wavepacket in the electronic initial state $F_1$. This results in distributions of $J$ states that are substantially different for the $F_1$ and $F_2$ states after the end of the pump-pulse interaction, as can be seen in the lower panel of Fig. 10. Tracing over the angular momentum states $J$ then results in a reduced density matrix with a low degree of coherence of only about 0.01 right after the pump-pulse interaction. Since the $J$ state distribution of states $F_1$ and $F_2$ is much narrower for interaction Hamiltonian D2, the dephasing is weaker and the coherence is not falling off as fast as compared to the interaction term D0.

The different roles of the interaction terms D0 and D2 can also be seen by directly looking at the density matrix. The temporal traces of the state multipoles $\mathrm{Real}(f^{11}_{K,0}(t))$, $\mathrm{Real}(f^{22}_{K,0}(t))$ and $\mathrm{Real}(f^{12}_{K,\pm 2}(t))$ for D0 and D2 are shown in Fig. 12. The traces are quite different for both cases. $\mathrm{Real}(f^{11}_{K,0}(t))$ for D0 recaptures the overall rotational dynamics of the full interaction Hamiltonian in Fig. 3. $\mathrm{Real}(f^{12}_{K,\pm 2}(t))$ for D2 recovers the fast modulation structures with the spin-orbit period of about 0.28 ps. Remarkably, all $\mathrm{Real}(f^{12}_{K,\pm 2}(t))$ for D0 are equal to zero. Note that $K = 2$ gives the largest contribution among $\mathrm{Real}(f^{12}_{K,\pm 2}(t))$ for D2.

For completeness, we show $W^{CC}(\theta,\beta,\tau)$ for $\beta = 0$ and $\beta = \pi/2$ for the interaction term D2 in Fig. 13. The modulation on the spin-orbit time scale is clearly visible and has almost flat angular dependence. This highlights the fact, that $K = 2$ is the only dominant component for D2. Since $d_{0,0}^{K=2}(\beta = \pi / 2) / d_{0,0}^{K=2}(\beta = 0) = -1/2$ , the contour plots for β = 0 and β = π/2 show a relative phase shift by exactly π.

The delay-dependent HHG signals of H9 and H15 for interaction terms D0 and D2 are shown in Fig. 14. The HHG spectra for D0 are entirely determined by the rotational wavepackets. The HHG signal following interaction with the term D0 is not sensitive to the electronic wavepacket. This can be seen from Eq. (36) and an analysis of the state multipoles of the density matrix. As can be seen in Eq. (36), the HHG cross channel is proportional to $\text{Real}(f_{K,\pm 2}^{12}(t))$. It turns out, however, that all $\text{Real}(f_{K,\pm 2}^{12}(t))$ are zero for the interaction Hamiltonian D0 (see Fig. 12). Note that the electronic wavepackets induced by D0 have non-vanishing state multipoles $\text{Real}(f_{K,\pm 1}^{12}(t))$ and $\text{Real}(f_{K,\pm 3}^{12}(t))$. Electronic wave-packets are excited, but the HHG process is not sensitive to those multipoles. The HHG traces for D2 is mainly dominated by the fast electronic modulation. When the polarization of the probe pulse changes from parallel to perpendicular with respect to the pump pulse, or β changes from 0 into π/2, the spectra for D2 show exactly a phase change of π, while the main structures of the spectra for D0 change roles (minima and maxima exchange). This is in accordance with the overall contributions $W^{DC}$ and $W^{CC}$ as plotted in Figs. 5 and 13.

## IV. Conclusions

We presented a derivation of a combined density matrix combined with the strong-field approximation and semiclassical quantitative rescattering approach, to quantitatively predict results of a novel HHG spectroscopic technique, to capture the combined rotational and electronic wavepackets prepared by impulsive Raman scattering on an ensemble of NO molecules. Our theoretical approach not only reproduces all qualitative features of the experiment, but also allows for a quantitative comparison of theory and experiment. Generally, the agreement between theory and experiment is good. Different interaction terms of the pump Hamiltonian were compared and studied in detail. The present formalism can be extended to other atomic or molecular systems for the study of the wavepacket dynamics with HHG. Our simulations support that HHG spectroscopy is a prospectively powerful probe mechanism for electronic and nuclear wavepackets.

**Acknowledgements**

We thank Prof. R. R. Lucchese for providing the photoionization matrix elements for the calculation of the HHG signals.

**Appendix A**

Substituting Eq. (27) into Eq. (30), we have

$$\begin{aligned}&D_{11}(\hat{R},t)=D_{22}(\hat{R},t)\\&=i\int d^3k\frac{1}{2}[\left\langle\pi^+\,|\,\boldsymbol{\mu}\cdot\hat{n}\,|\,\mathbf{k}\right\rangle\int_0^t dt'\left\langle\mathbf{k}'\,|\,\boldsymbol{\mu}\cdot\hat{\boldsymbol{\varepsilon}}_2(t')\,|\,\pi^+\right\rangle\\&+\left\langle\pi^-\,|\,\boldsymbol{\mu}\cdot\hat{n}\,|\,\mathbf{k}\right\rangle\int_0^t dt'\left\langle\mathbf{k}'\,|\,\boldsymbol{\mu}\cdot\hat{\boldsymbol{\varepsilon}}_2(t')\,|\,\pi^-\right\rangle]e^{-iS(t,t')}+c.c\end{aligned},$$

(A.1)

and

$$\begin{aligned}&D_{12}(\hat{R},t)=D_{21}(\hat{R},t)^*\\&=i\int d^3k\frac{1}{2}[\left\langle\pi^+\,|\,\boldsymbol{\mu}\cdot\hat{n}\,|\,\mathbf{k}\right\rangle\int_0^t dt'\left\langle\mathbf{k}'\,|\,\boldsymbol{\mu}\cdot\hat{\boldsymbol{\varepsilon}}_2(t')\,|\,\pi^-\right\rangle\\&+\left\langle\pi^-\,|\,\boldsymbol{\mu}\cdot\hat{n}\,|\,\mathbf{k}\right\rangle\int_0^t dt'\left\langle\mathbf{k}'\,|\,\boldsymbol{\mu}\cdot\hat{\boldsymbol{\varepsilon}}_2(t')\,|\,\pi^+\right\rangle]e^{-iS(t,t')}+c.c\end{aligned}. \qquad (A.2)$$

The explicit coordinate representation of the orbitals $\pi^+$ and $\pi^-$ on $\chi$ is given by $-e^{-i\chi}$ and $e^{i\chi}$ respectively [43]. The $\chi$ dependent phase factor in Eq. (A.1) cancel each other, since the matrix elements are of bra/ket combination $\left\langle\pi^+|\cdots|\pi^+\right\rangle$ and $\left\langle\pi^-|\cdots|\pi^-\right\rangle$. In Eq. (A.2), $\left\langle\pi^+|\cdots|\pi^-\right\rangle$ and $\left\langle\pi^-|\cdots|\pi^+\right\rangle$ result in the phase factor $e^{i2\chi}$ and $e^{-i2\chi}$, respectively. In practice, due to the spin-orbit interaction the radial wavefunction pertaining to the states $\pi^+$ and $\pi^-$ are not identical, and the spin-orbit splitting lifts the degeneracy. The reduced radial matrix elements in Eq. (A.1) and Eq. (A.2) will however vary only very little by inclusion of the spin-orbit interaction. We therefore suppose that the radial wavefunction for both $\pi^+$ and $\pi^-$ states are given by a single orbital, $\Phi_{\text{HOMO}}$. In practice this orbital can be the highest-occupied molecular orbital from an electronic structure calculation, or a Dyson orbital, which would be more appropriate, when deriving the strong-field approximation starting from a many-body wavefunction. In the above equations we can therefore factor out the dependence on the angle $\chi$, since the product $\boldsymbol{\mu}\cdot\hat{n}$ and $\boldsymbol{\mu}\cdot\hat{\boldsymbol{\varepsilon}}_2$ do not depend on that angle explicitly. We get

$$\begin{aligned} &D_{11}(\hat{R},t)=D_{22}(\hat{R},t) \\ &\simeq i\int d^3k\left\langle\Phi_{\mathrm{HOMO}}(\theta)\,|\,\boldsymbol{\mu}\cdot\hat{n}\,|\,\mathbf{k}\right\rangle\int_0^t dt'\left\langle\mathbf{k}'\,|\,\boldsymbol{\mu}\cdot\hat{\boldsymbol{\varepsilon}}_2(t')\,|\,\Phi_{\mathrm{HOMO}}(\theta)\right\rangle e^{-iS(t,t')}+c.c \\ &=i\int d^3k\left\langle\Phi_{\mathrm{HOMO}}(\theta,\chi=0)\,|\,\boldsymbol{\mu}\cdot\hat{n}\,|\,\mathbf{k}\right\rangle\int_0^t dt'\left\langle\mathbf{k}'\,|\,\boldsymbol{\mu}\cdot\hat{\boldsymbol{\varepsilon}}_2(t')\,|\,\Phi_{\mathrm{HOMO}}(\theta,\chi=0)\right\rangle e^{-iS(t,t')}+c.c \end{aligned}\quad, \tag{A.3}$$

and

$$\begin{aligned} &D_{12}(\hat{R},t)=D_{21}(\hat{R},t)^* \\ &\simeq i\int d^3k\left\langle\Phi_{\mathrm{HOMO}}(\theta,\chi=0)\,|\,\boldsymbol{\mu}\cdot\hat{n}\,|\,\mathbf{k}\right\rangle\int_0^t dt'\left\langle\mathbf{k}'\,|\,\boldsymbol{\mu}\cdot\hat{\boldsymbol{\varepsilon}}_2(t')\,|\,\Phi_{\mathrm{HOMO}}(\theta,\chi=0)\right\rangle e^{-iS(t,t')}\frac{e^{-i2\chi}+e^{i2\chi}}{2}+c.c \end{aligned}\,. \tag{A.4}$$

We can rewrite Eq. (A.3) and Eq. (A.4) as

$$\begin{aligned} &D_{\Xi\Xi'}(\hat{R},t) \\ &=i\int d^3k\left\langle\Phi_{\mathrm{HOMO}}(\theta,\chi=0)\,|\,\boldsymbol{\mu}\cdot\hat{n}\,|\,\mathbf{k}\right\rangle \\ &\times\int_0^t dt'\left\langle\mathbf{k}'\,|\,\boldsymbol{\mu}\cdot\hat{\boldsymbol{\varepsilon}}_2(t')\,|\,\Phi_{\mathrm{HOMO}}(\theta,\chi=0)\right\rangle e^{-iS(t,t')}\frac{e^{-i2(\Xi-\Xi')\chi}+e^{i2(\Xi-\Xi')\chi}}{2}+c.c \end{aligned}\,. \tag{A.5}$$

## Appendix B

The amplitude for photoionization can be written as [50]

$$I_{\hat{k},\hat{n}}^{\Lambda_i,\Lambda_f}=\sum_{l,m,\mu}\left\langle\Psi^{\Lambda_i}\,|\,r_\mu\,|\,\Phi^{\Lambda_f}\psi_{klm}\right\rangle Y_{l,m}^*(\hat{k})Y_{1,\mu}^*(\hat{n}), \tag{A.6}$$

where $\Lambda_i$ is the quantum number of the orbital angular momentum projected on the molecular axis, *i.e.*, the azimuthal angular momentum, in the initial state $\left|\Psi^{\Lambda_i}\right\rangle$, $\Lambda_f$ is the corresponding azimuthal quantum number of the final ionic state $\left|\Phi^{\Lambda_f}\right\rangle$, $\left|\psi_{klm}\right\rangle$ is the partial wavefunction of the continuum electron. $\hat{k}$ and $\hat{n}$ are the directions of emission of the photoelectron and direction of polarization of the linearly polarized light, respectively. In the case of ionization of NO to its ground ionic state, $\Lambda_i=\pm1$ and $\Lambda_f=0$. And $\left|\Psi^{\Lambda_i=\pm1}\right\rangle$ correspond to $\pi^+$ and $\pi^-$ states. The non-zero matrix elements follow the selection rule

$$\Lambda_i=\mu+\Lambda_f+m=\mu+m\,. \tag{A.7}$$

Note that HHG is the inverse process of photoionization. In the HHG process, $\hat{k}$ is supposed to be parallel to the driving laser, and $\hat{n}$ is the direction of the emitted HHG spectrum. And the case of $\hat{k}\parallel\hat{n}$ would contribute most to the recombination matrix [37]. So both vectors $\hat{k}$ and $\hat{n}$ can be replaced with the angles $(\theta,\chi)$ with respect to the molecular axis, and Eq. (A.6) can be

rewritten as

$$
\begin{aligned}
I_{\hat{k},\hat{n}}^{\Lambda_i,\Lambda_f}(\theta,\chi) &= \sum_{l,m,\mu} \left\langle \Psi^{\Lambda_i} \mid r_\mu \mid \Phi^{\Lambda_f}\psi_{klm} \right\rangle (-1)^m \sqrt{\frac{2l+1}{4\pi}\frac{(l-m)!}{(l+m)!}} P_l^m(\cos\theta) e^{-im\chi} \\
&\quad \times (-1)^\mu \sqrt{\frac{2+1}{4\pi}\frac{(1-\mu)!}{(1+\mu)!}} P_1^\mu(\cos\theta) e^{-i\mu\chi} \\
&= \sum_{l,m,\mu} \left\langle \Psi^{\Lambda_i} \mid r_\mu \mid \Phi^{\Lambda_f}\psi_{klm} \right\rangle \frac{\sqrt{3(2l+1)}}{4\pi} \sqrt{\frac{(l-m)!(1-\mu)!}{(l+m)!(1+\mu)!}} P_l^m(\cos\theta) P_1^\mu(\cos\theta)(-1)^{m+\mu} e^{-i(m+\mu)\chi} \\
&= (-1)^{\Lambda_i} e^{-i\Lambda_i\chi} \sum_{l,m,\mu} \left\langle \Psi^{\Lambda_i} \mid r_\mu \mid \Phi^{\Lambda_f}\psi_{klm} \right\rangle \frac{\sqrt{3(2l+1)}}{4\pi} \sqrt{\frac{(l-m)!(1-\mu)!}{(l+m)!(1+\mu)!}} P_l^m(\cos\theta) P_1^\mu(\cos\theta) \\
&= (-1)^{\Lambda_i} e^{-i\Lambda_i\chi} I_{\hat{k},\hat{n}}^{\Lambda_i,\Lambda_f}(\theta,\chi=0)
\end{aligned}
\tag{A.8}
$$

Eq. (A.8) expresses that a $\chi$ dependent phase term in the ionization matrix element can be factored for the expression of the photoionization dipole matrix element. We therefore can write the ionization matrix from $|\Psi^{\Lambda_i}>$ state ($d_{ion}^{\Lambda_i}(\theta,\chi)$) and recombination matrix to $|\Psi^{\Lambda_f}>$ state ($d_{rec}^{\Lambda_f}(\theta,\chi)$) as

$$
\begin{cases}
d_{ion}^{\Lambda_i}(\theta,\chi) = (-1)^{\Lambda_i} e^{-i\Lambda_i\chi} d_{ion}^{\Lambda_i}(\theta,\chi=0) \\
d_{rec}^{\Lambda_f}(\theta,\chi) = (-1)^{\Lambda_f} e^{i\Lambda_f\chi} d_{rec}^{\Lambda_f}(\theta,\chi=0)
\end{cases},
\tag{A.9}
$$

Note that $d_{ion}^{\Lambda_i}(\theta,\chi=0) = d_{ion}^{\Lambda_f}(\theta,\chi=0)$ and $d_{rec}^{\Lambda_i}(\theta,\chi=0) = d_{rec}^{\Lambda_f}(\theta,\chi=0)$. So Eq. (A.1) and Eq. (A.2) can be rewritten as

$$
\begin{aligned}
D_{11}(\hat{R},t) &= D_{22}(\hat{R},t) \\
&= \frac{1}{2}[d_{rec}^{\Lambda_f=1}(\theta,\chi) d_{ion}^{\Lambda_i=1}(\theta,\chi) + d_{rec}^{\Lambda_f=-1}(\theta,\chi) d_{ion}^{\Lambda_i=-1}(\theta,\chi)] a_{ewp}, \\
&= d_{rec}^{\Lambda_f=1}(\theta,\chi=0) a_{ewp} d_{ion}^{\Lambda_i=1}(\theta,\chi=0)
\end{aligned}
\tag{A.10}
$$

and

$$
\begin{aligned}
D_{12}(\hat{R},t) &= D_{21}(\hat{R},t)^* \\
&= \frac{1}{2}[d_{rec}^{\Lambda_f=1}(\theta,\chi) d_{ion}^{\Lambda_i=-1}(\theta,\chi) + d_{rec}^{\Lambda_f=-1}(\theta,\chi) d_{ion}^{\Lambda_i=1}(\theta,\chi)] a_{ewp}. \\
&= d_{rec}^{\Lambda_f=1}(\theta,\chi=0) a_{ewp} d_{ion}^{\Lambda_i=-1}(\theta,\chi=0) \frac{e^{-i2\chi}+e^{i2\chi}}{2}
\end{aligned}
\tag{A.11}
$$

We can rewrite Eq. (A.10) and Eq. (A.11) as

$$
D_{\Xi\Xi'}(\hat{R},t) = d_{rec}^{\Lambda_\Xi}(\theta,\chi=0) a_{ewp} d_{ion}^{\Lambda_{\Xi'}}(\theta,\chi=0) \frac{e^{-i2(\Xi-\Xi')\chi}+e^{i2(\Xi-\Xi')\chi}}{2}.
\tag{A.12}
$$

## Figure Captions

Fig. 1 (Color online) Panel (a): Illustration of HHG starting from a coherent superposition of two electronic states: the first two pathways, HHG from different electronic channels, happen independently (direct HHG channels); the other two pathways are connecting different electronic states (cross channel HHG). These channels only contribute to a macroscopic signal when they are coherently connected. In the density matrix description it is the off-diagonal matrix element (the coherence) that determines the HHG signal; Panel (b) rotational level diagram of the two electronic states $F_1$ and $F_2$ of NO Fig. 2 (Color online) $J$ state populations for states $F_1$ and $F_2$ before and after the alignment pulse with the inset of the temporal populations for those two states

Fig. 3 (Color online) Temporal variations of the multipole coefficients $\mathrm{Real}(f_{K,0}^{11}(t))$ and $\mathrm{Real}(f_{K,0}^{22}(t))$ for $K$= 0, 2, 4 and 6, that are related to the electronic occupations

Fig. 4 (Color online) Temporal variations of the multipole coefficients $\mathrm{Real}(f_{K,\pm 2}^{12}(t))$ for $K$= 2, 4 and 6, that are related to the electronic coherence

Fig. 5 (Color online) Temporal contour plots of the quantity $W^{DC}(\theta,\beta,\tau)$ for $\beta = 0$ and $\pi/2$, which corresponds to parallel and perpendicular polarizations of the pump and probe pulses, respectively

Fig. 6 (Color online) Temporal contour plots of the quantity $W^{CC}(\theta,\beta,\tau)$ for $\beta = 0$ and $\pi/2$.

Fig. 7 (Color online) Yield of the 9$^{th}$ harmonic (H9) as a function of pump-probe time delay for $\beta = 0$ and the Fourier transform for both theory (left panel) and experiment (right panel). Furthermore, we show the expected signals (shifted up) by omitting the coherent cross channel contribution

Fig. 8 (Color online) Yield of the 15$^{th}$ harmonic (H15) as a function of pump-probe time delay for $\beta = 0$ and the Fourier transform for both theory (left panel) and experiment (right panel). Furthermore, we show the expected signals (shifted up) by omitting the coherent cross channel contribution

Fig. 9 (Color online) Yield of harmonics H9 and H15 as a function of pump-probe delay time for parallel ($\beta = 0$) and perpendicular ($\beta =\pi/2$) polarization of pump and probe-laser fields

Fig. 10 (Color online) $J$ state populations for states $F_1$ and $F_2$ before and after the alignment pulse. The inset shows the temporal evolution of the total occupation in states $F_1$ and $F_2$. The upper panel shows results for including only the first term (D0) of the interaction Hamiltonian of Eq. (7), the lower panel shows results for treating only the second term (D2)

Fig. 11 (Color online) Temporal evolution of the off-diagonal reduced electronic density matrix element $\hat{\rho}_{12}^{el}(t)$ and the degree of coherence $\mathbb{C}(t)$ for interactions D0, D2 and D0+D2 terms of interaction Hamiltonian of Eq. (7)

Fig. 12 (Color online) Temporal evolution of the multipole coefficients Real($f^{11}_{K,0}(t)$) and Real($f^{22}_{K,0}(t)$) for $K$= 0, 2, 4 and 6, and Real($f^{12}_{K,\pm2}(t)$) for $K$= 2, 4 and 6 for interactions D0 and D2

Fig. 13 (Color online) Temporal evolution of the purely geometric quantity $W^{CC}(\theta,\beta,\tau)$ for interaction D2 for β = 0 and β = π/2

Fig. 14 (Color online) Yield of harmonics H9 and H15 as a function of the pump-probe time delay for the interactions D0 and D2 for β = 0 and π/2

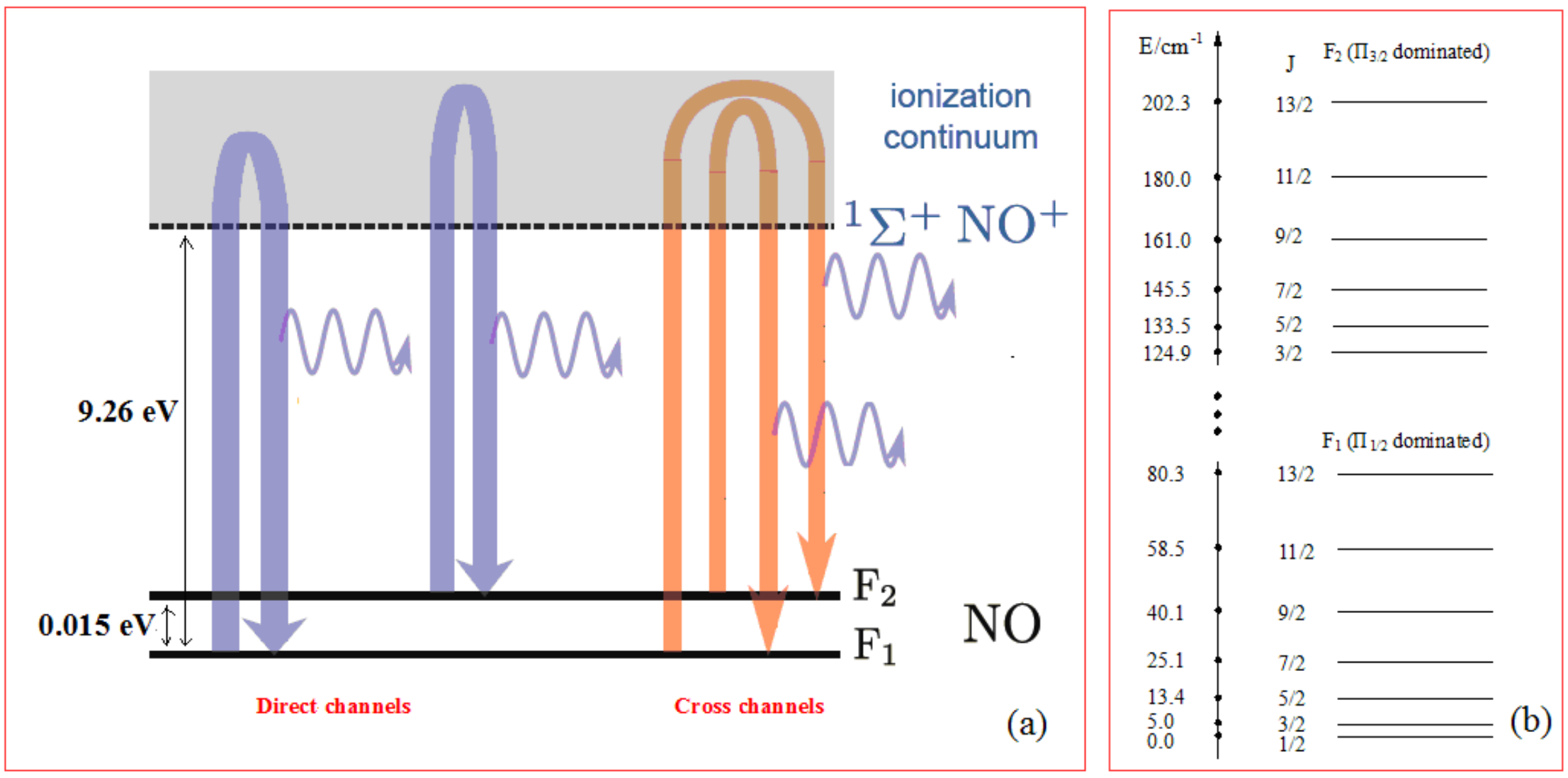

Fig. 1

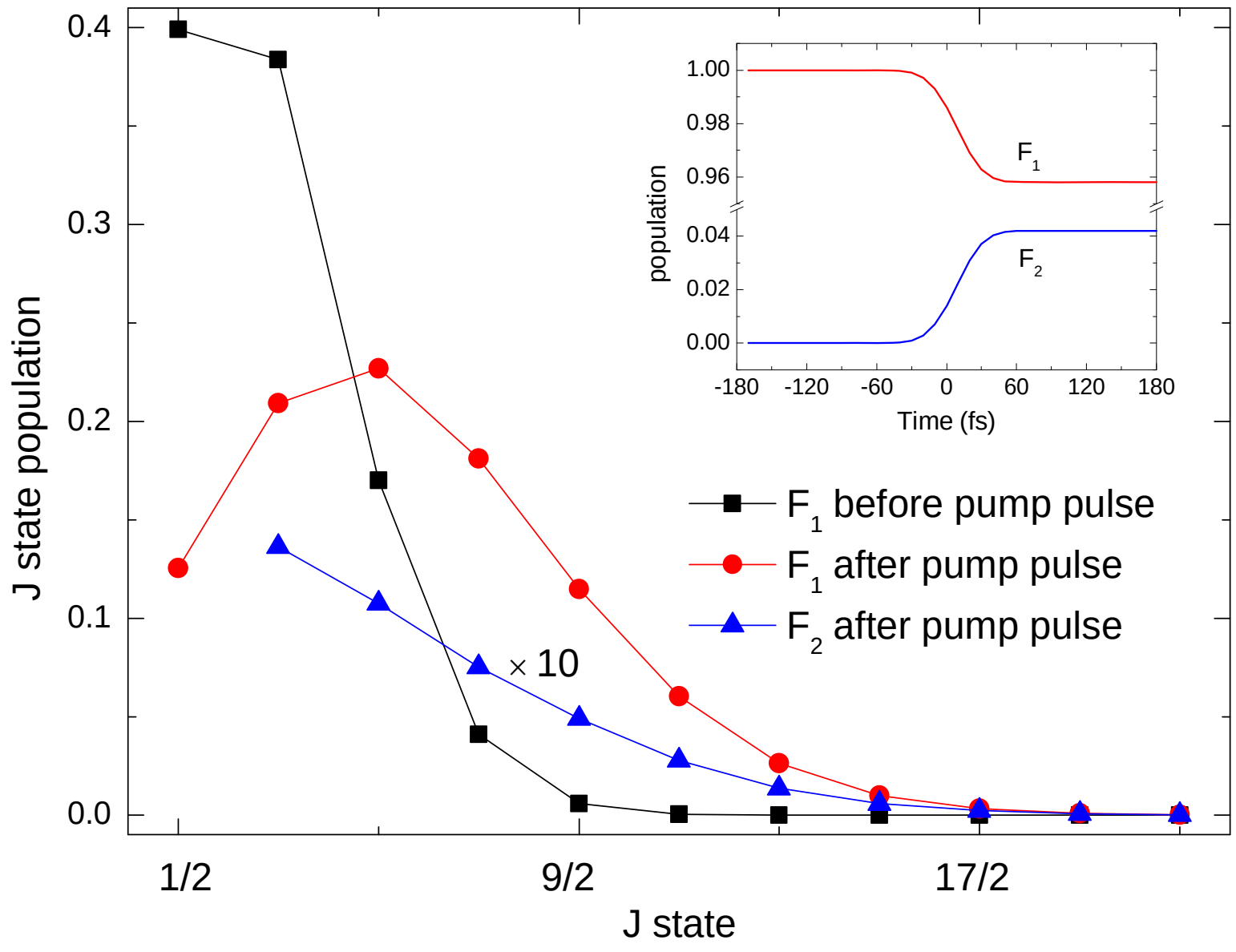

Fig. 2

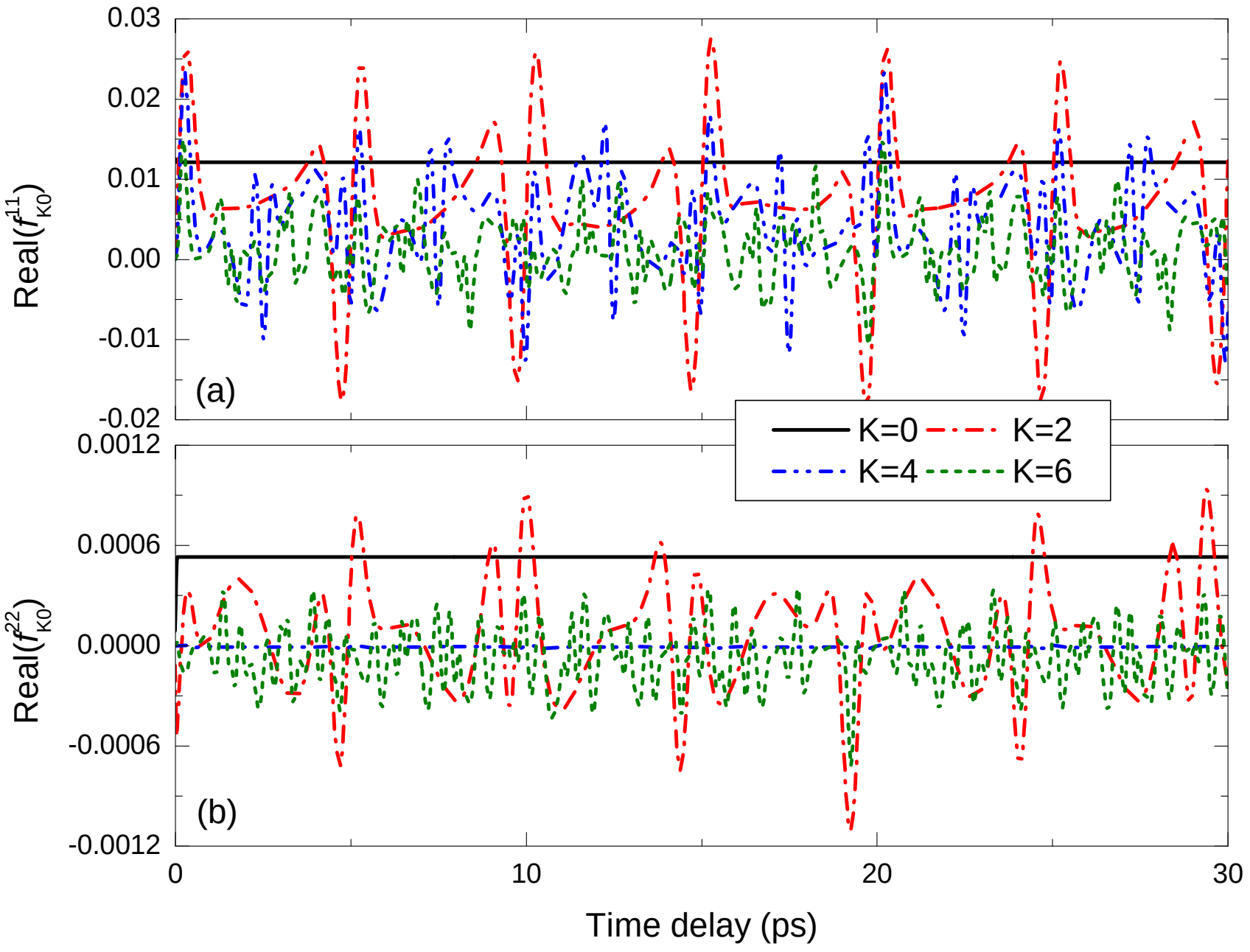

Fig. 3

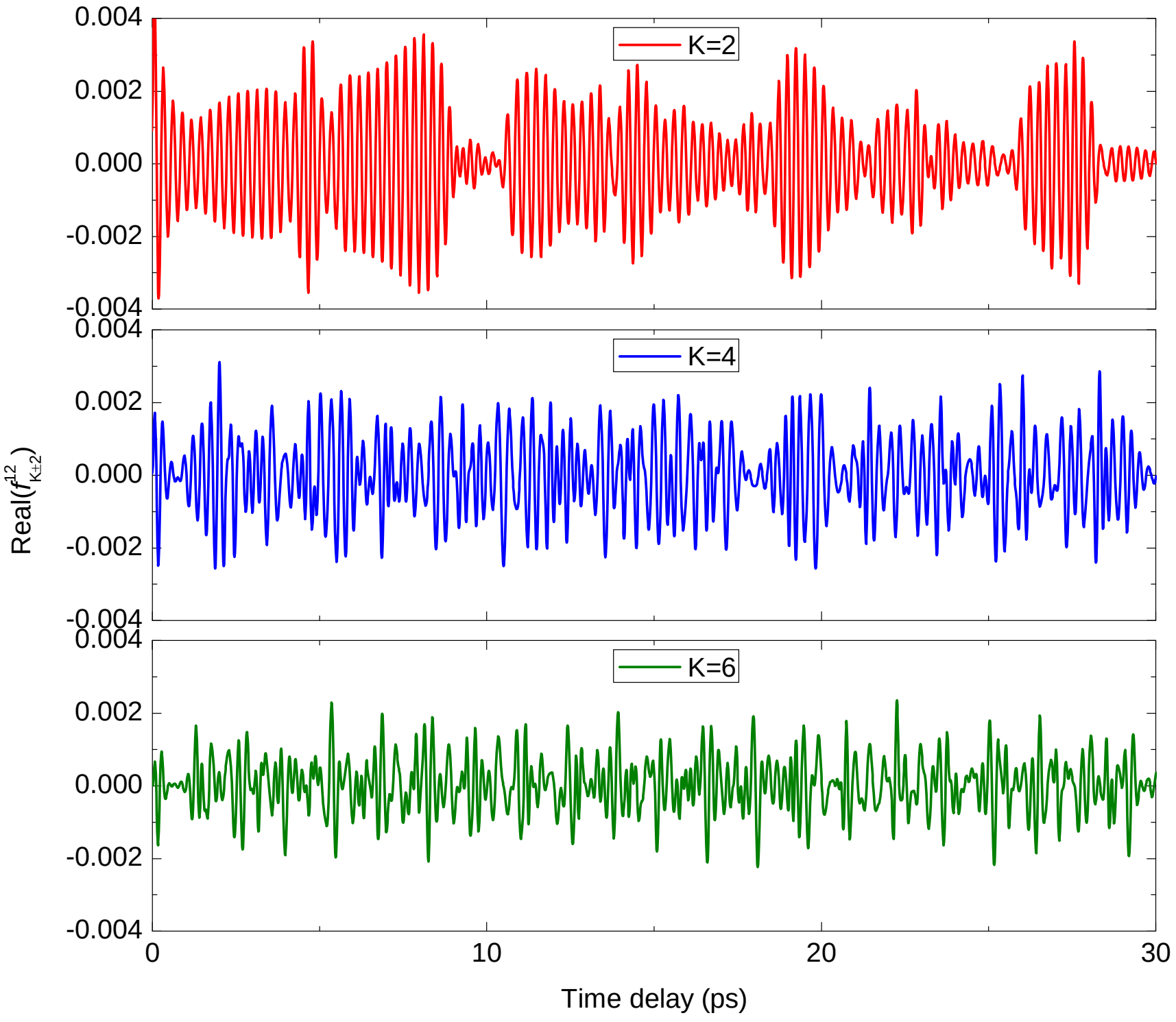

Fig. 4

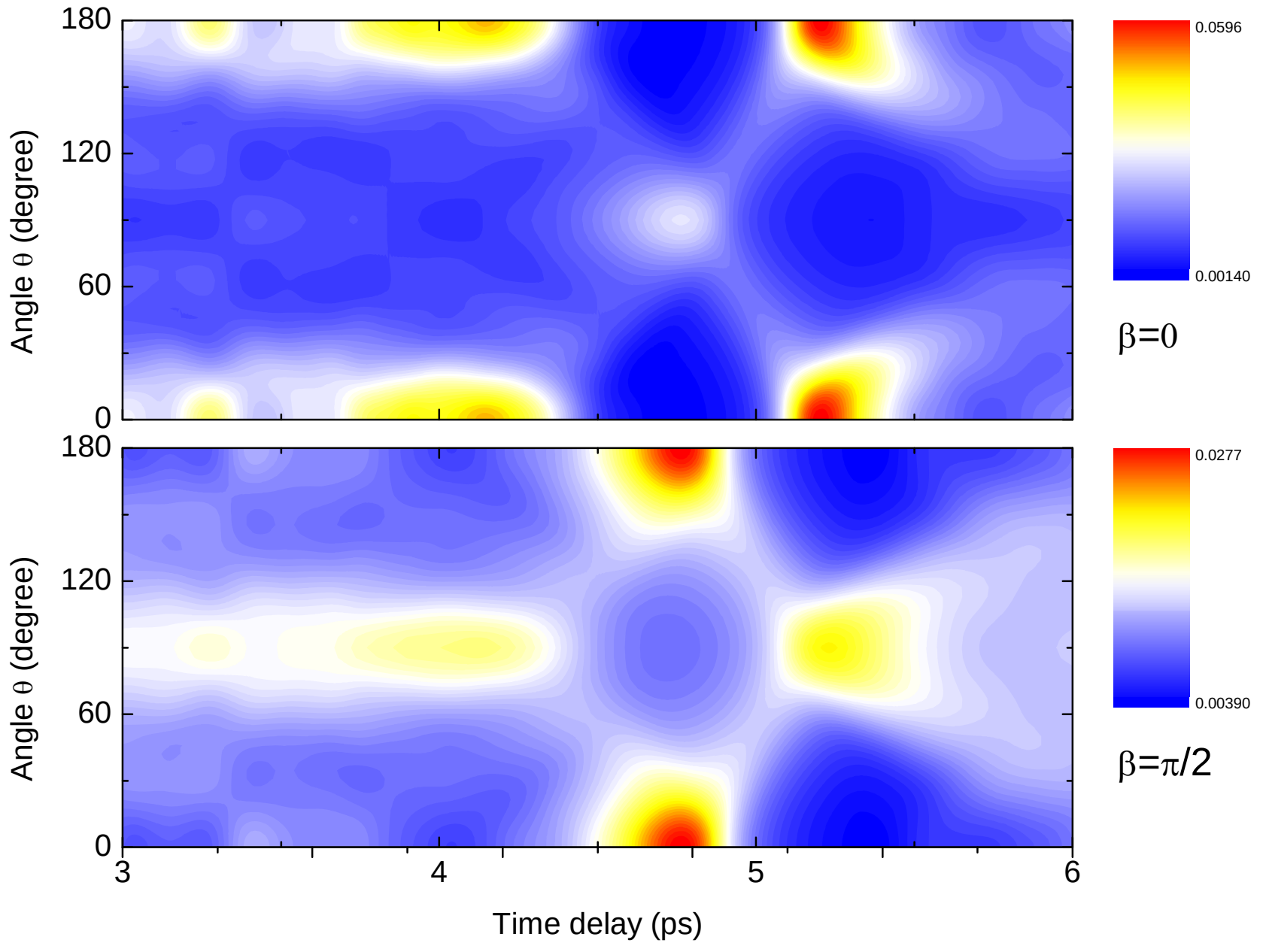

Fig. 5

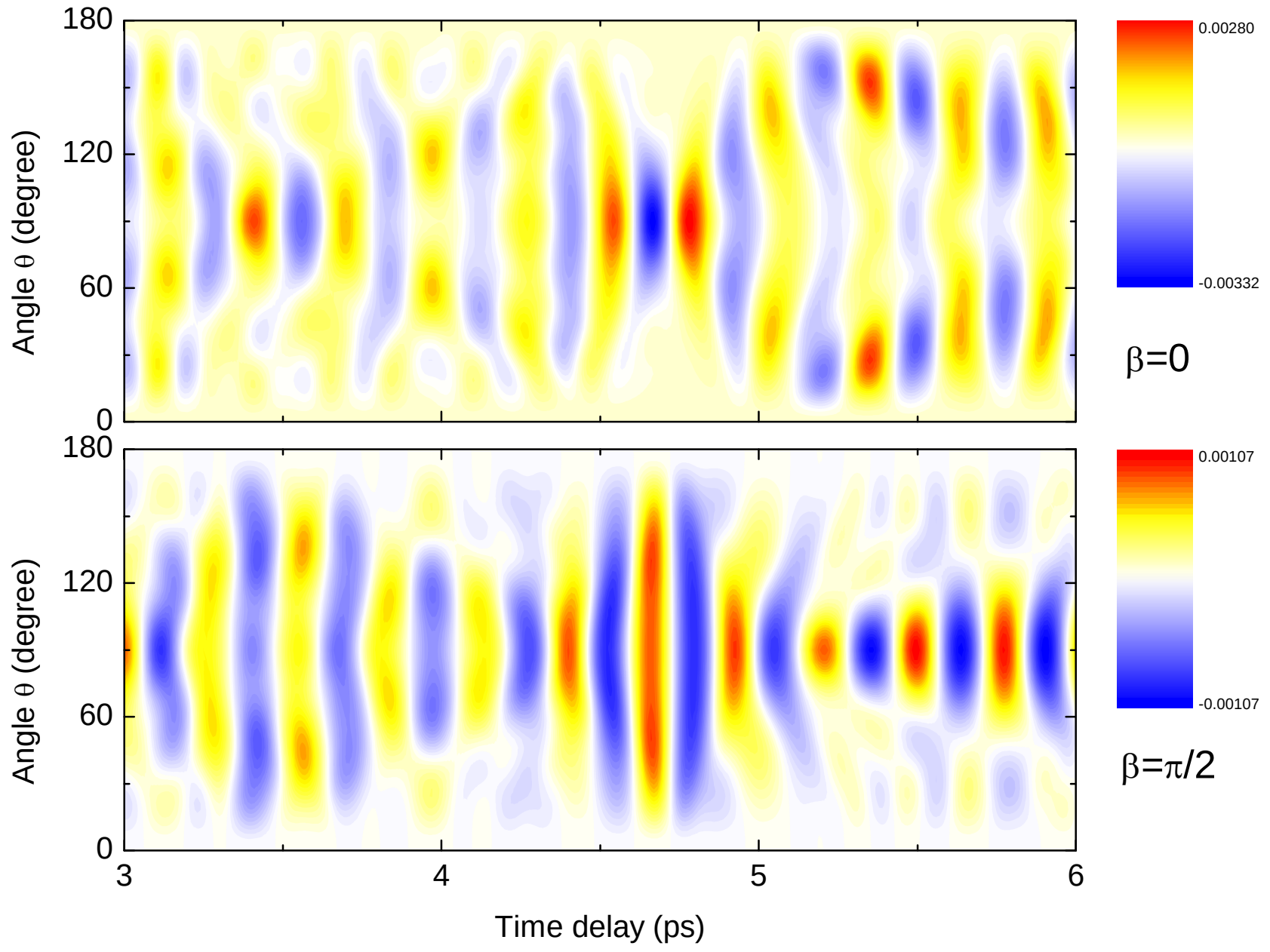

Fig. 6

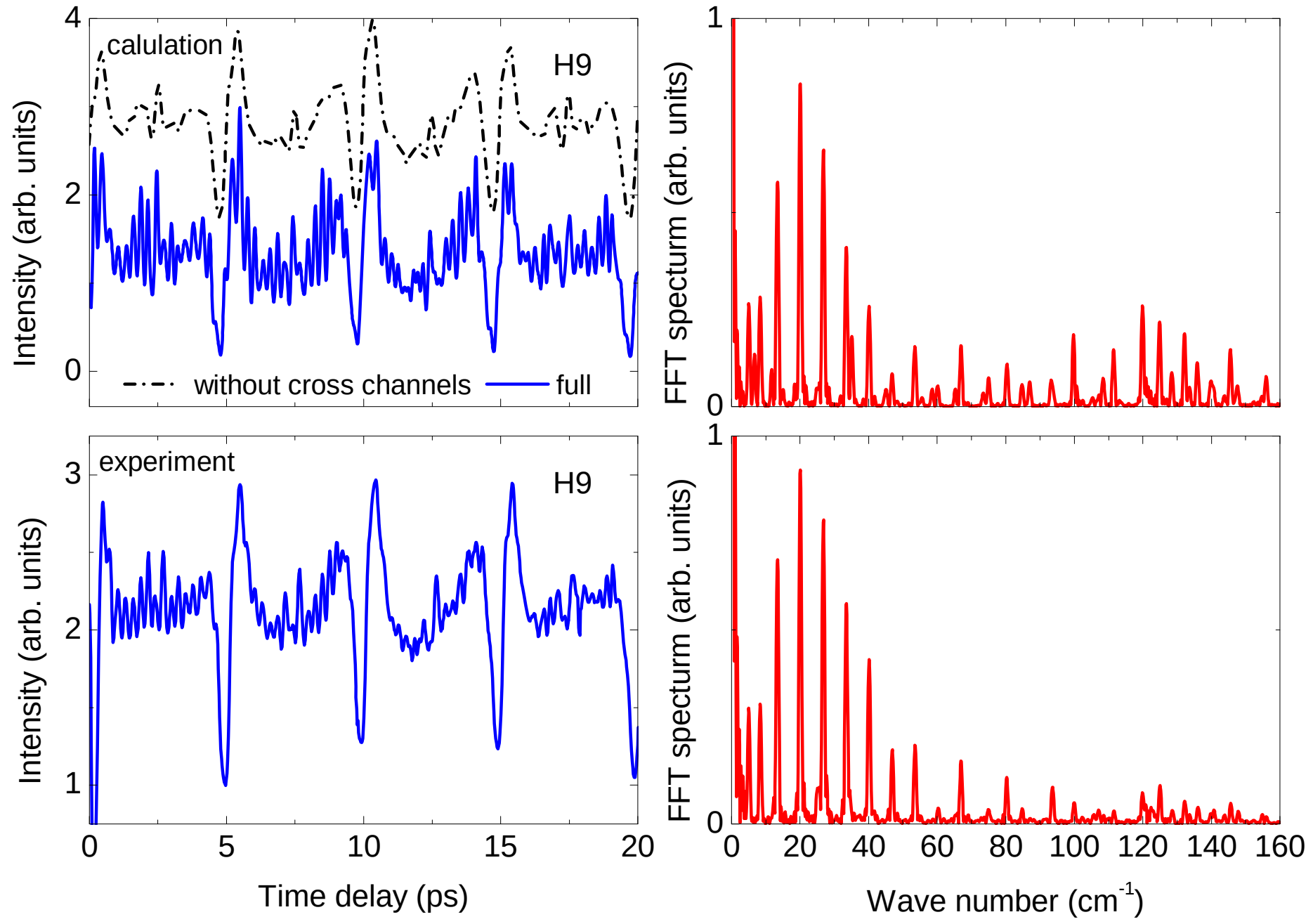

Fig. 7

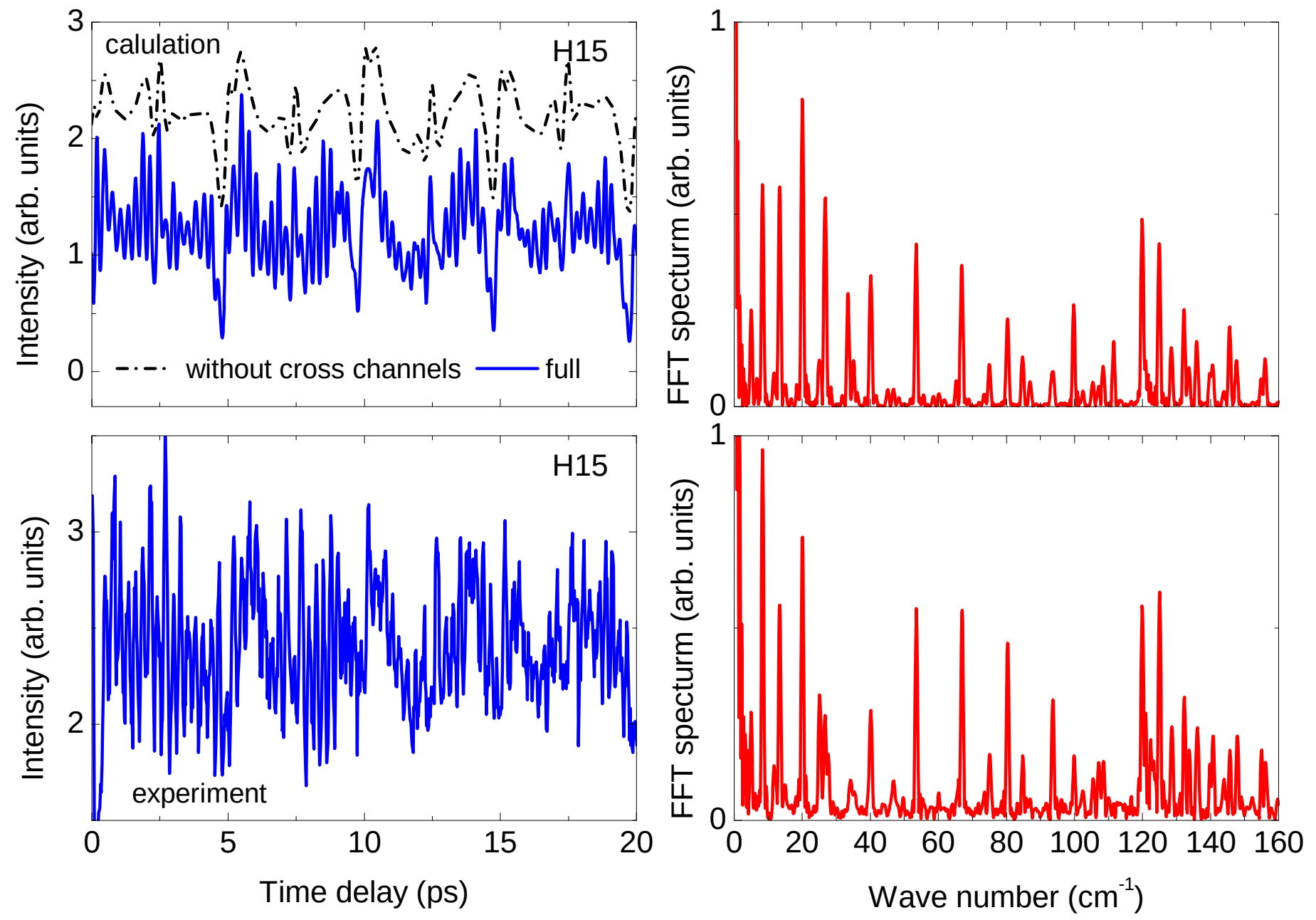

Fig. 8

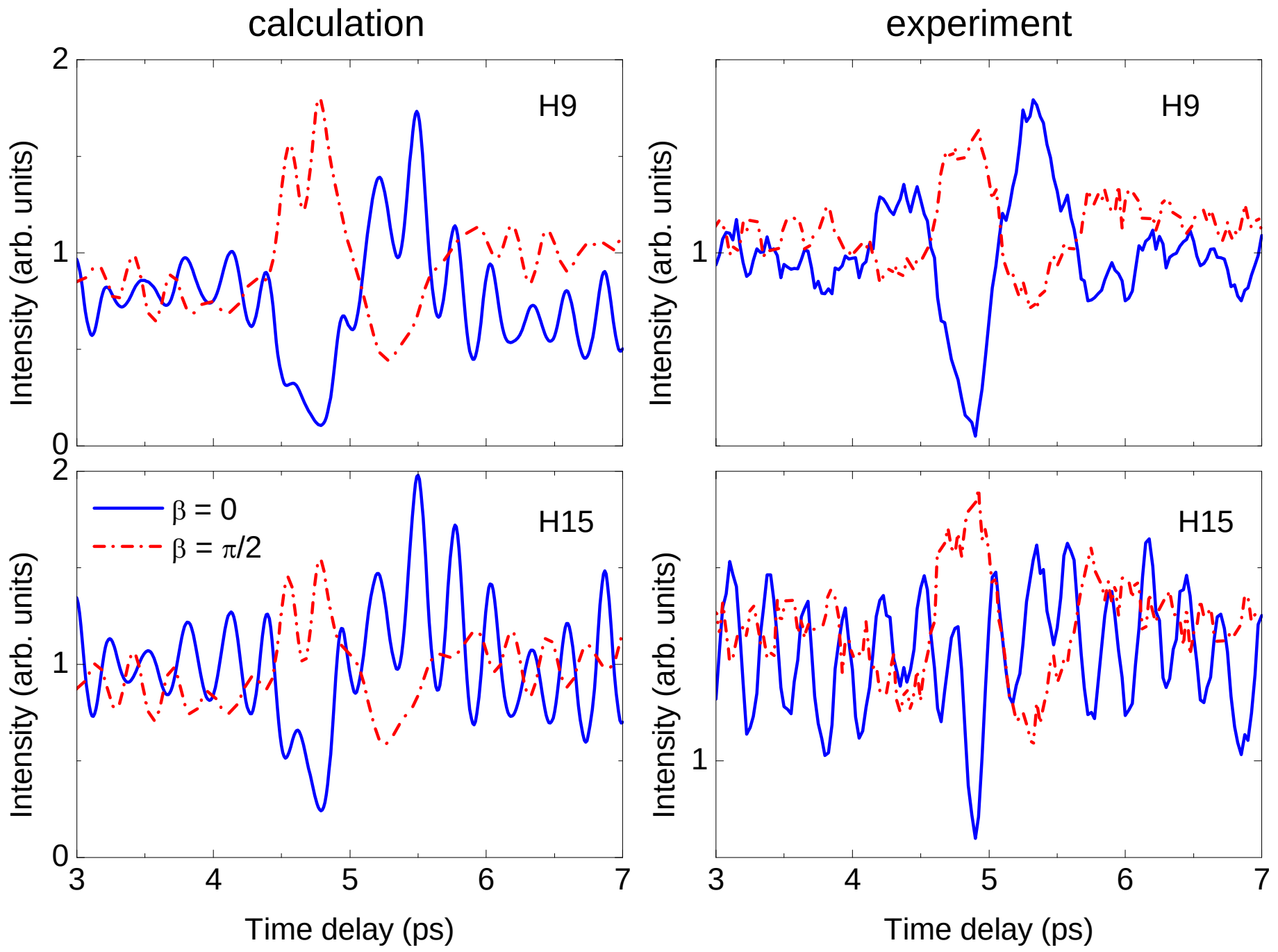

Fig. 9

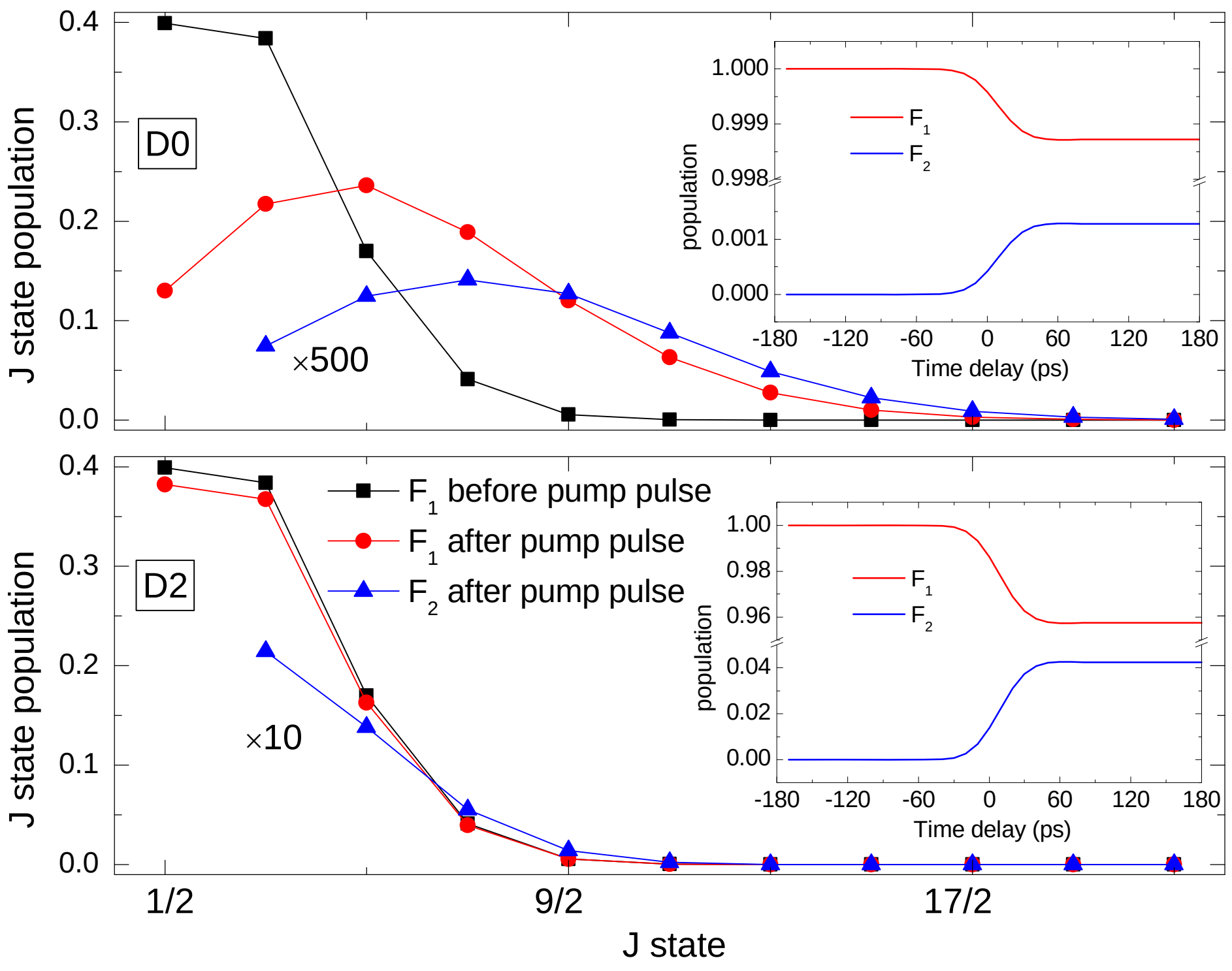
D0
D2
×500
×10
J state population
J state
F1 before pump pulse
F1 after pump pulse
F2 after pump pulse
population
Time delay (ps)

Fig .10

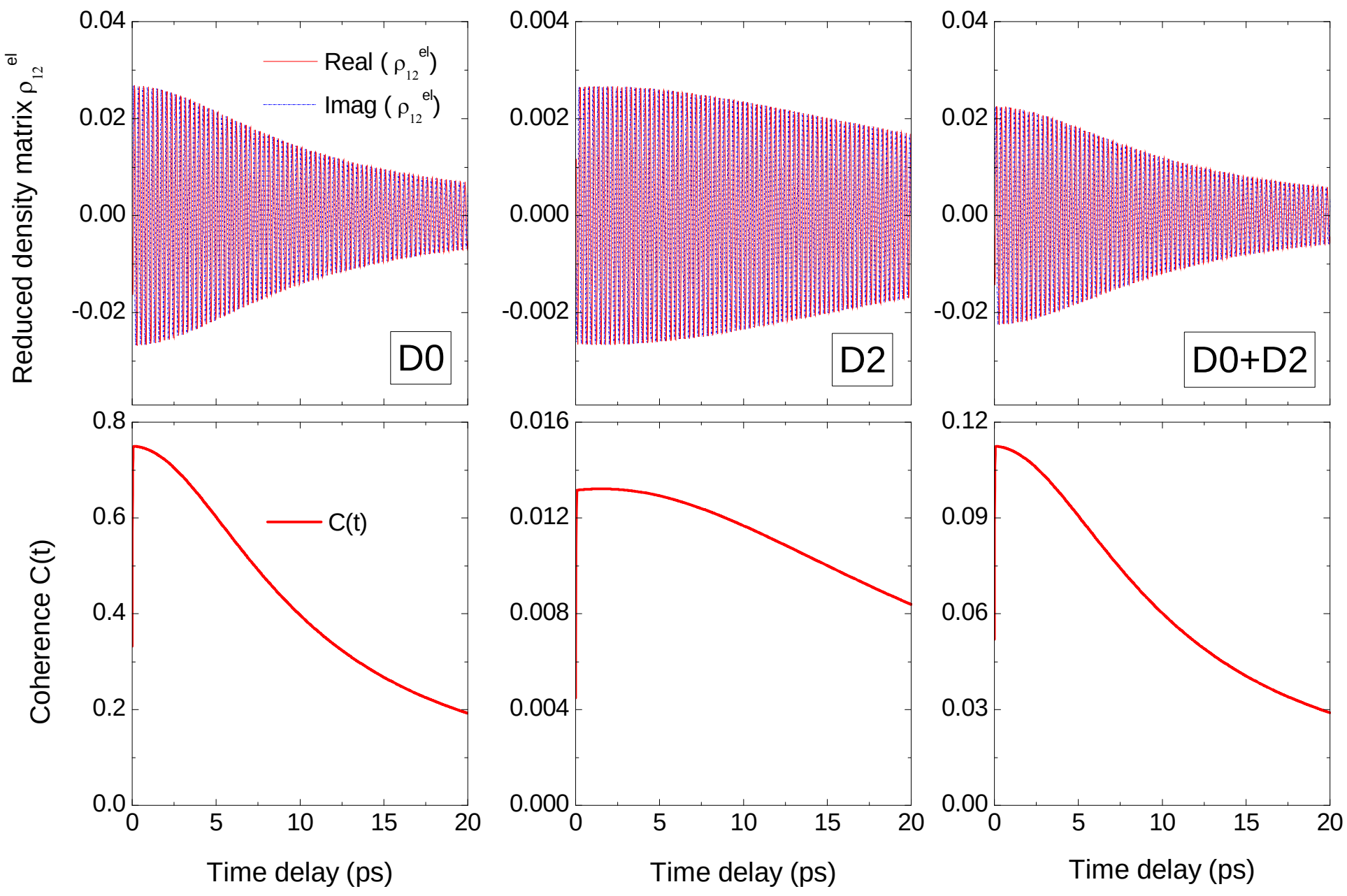

Fig. 11

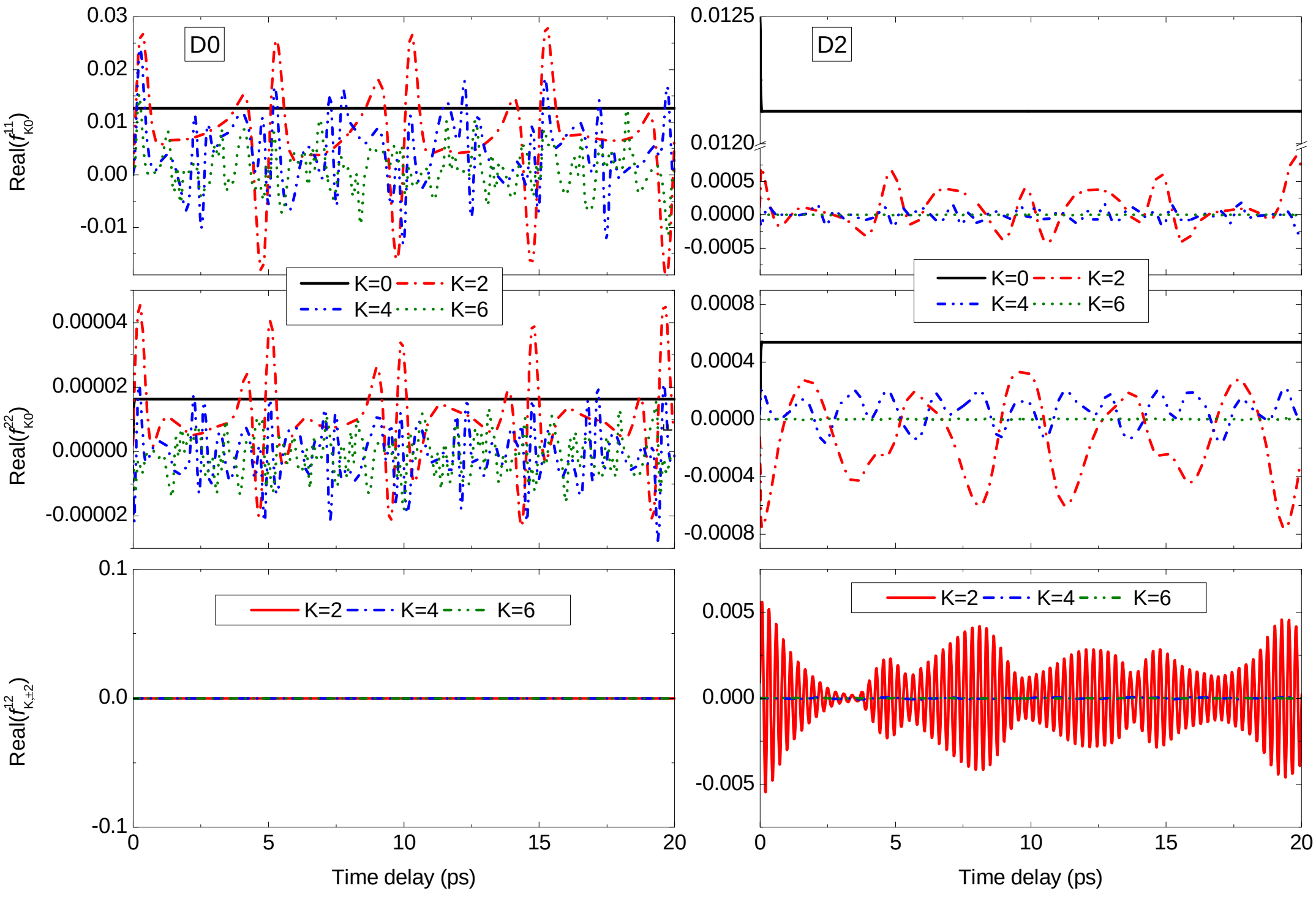

Fig. 12

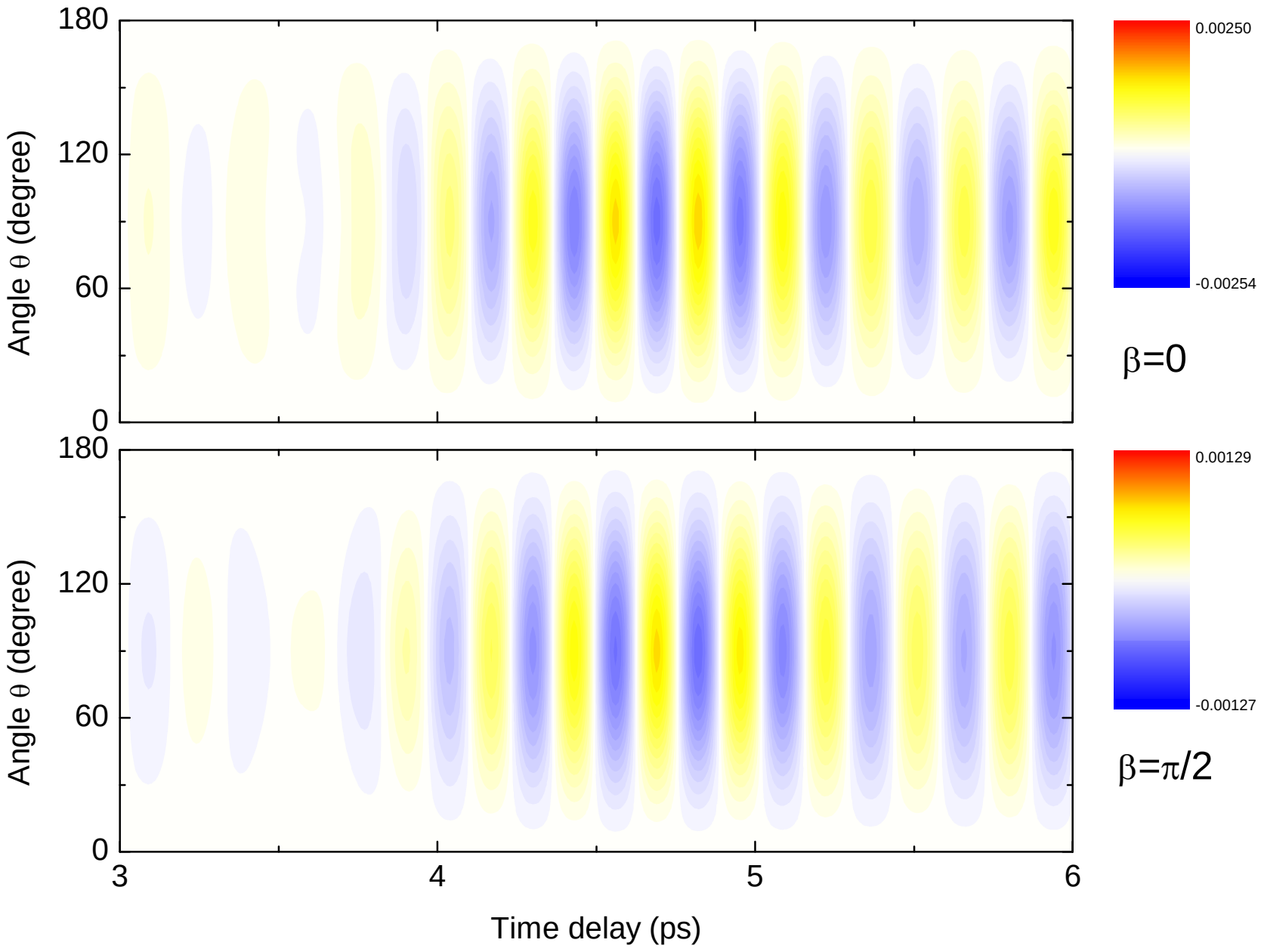

Fig. 13

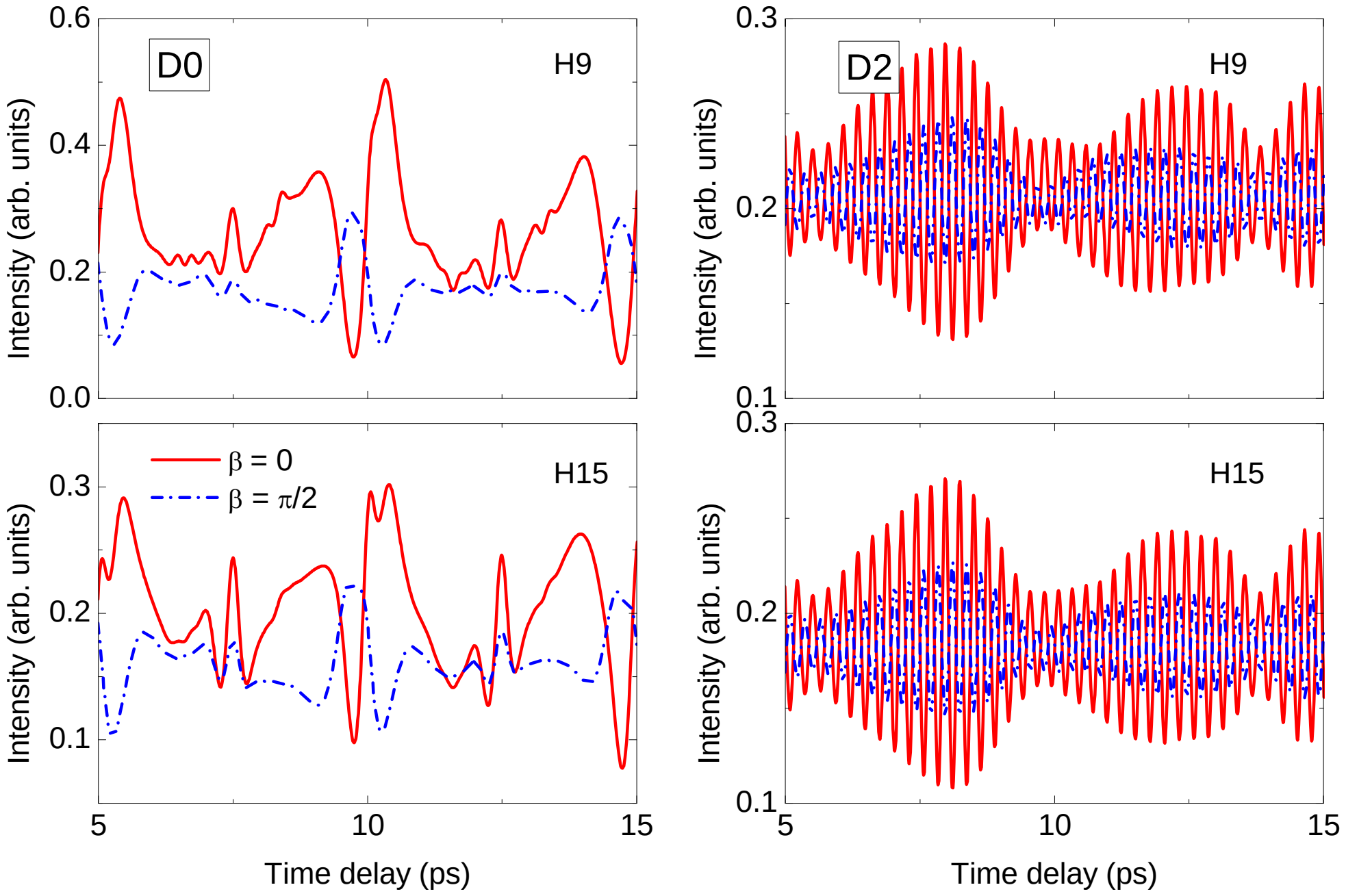

Fig. 14